\documentclass[pdflatex,sn-nature]{sn-jnl}

\usepackage{graphicx}%
\usepackage{multirow}%
\usepackage{amsmath,amssymb,amsfonts}%
\usepackage{amsthm}%
\usepackage[title]{appendix}%
\usepackage{xcolor}%
\usepackage{textcomp}%
\usepackage{manyfoot}%
\usepackage{booktabs}%
\usepackage{algorithm}%
\usepackage{algorithmicx}%
\usepackage{algpseudocode}%
\usepackage{listings}%
\usepackage{multicol}
\usepackage{setspace}
\usepackage{lineno} 
\usepackage{braket}
\usepackage{hyperref}
\hypersetup{colorlinks,linkcolor={blue},citecolor={blue},urlcolor={red}}

\theoremstyle{thmstyleone}

\theoremstyle{thmstyletwo}%

\theoremstyle{thmstylethree}%

\begin{document}

\title[Article Title]{An integrated all-van der Waals nanobeam laser}

\author*[1]{\fnm{Aris} \sur{Koulas-Simos}}\email{aris.koulas-simos@tu-berlin.de}

\author[2]{\fnm{Pietro} \sur{Metuh}}\email{piemet@dtu.dk}

\author[2]{\fnm{Athanasios} \sur{Paralikis}}\email{athpa@dtu.dk}

\author[1]{\fnm{Kartik} \sur{Gaur}}\email{kartik.gaur@tu-berlin.de}

\author[1]{\fnm{Maximilian} \sur{Klonz}}\email{m.klonz@tu-berlin.de}

\author[1]{\fnm{Imad} \sur{Limame}}\email{imad.limame@tu-berlin.de}

\author[1,3]{\fnm{B\'arbara L. T.} \sur{Rosa}}\email{bltr@unicamp.br}

\author[1]{\fnm{Chirag C.} \sur{Palekar}}\email{c.palekar@tu-berlin.de}

\author*[2]{\fnm{Battulga} \sur{Munkhbat}}\email{bamunk@dtu.dk}

\author*[1]{\fnm{Stephan} \sur{Reitzenstein}}\email{stephan.reitzenstein@tu-berlin.de}

\affil[1]{\orgdiv{Institute for Physics and Astronomy}, \orgname{Technical University of Berlin}, \orgaddress{\street{Hardenbergstr. 36}, \postcode{10623} \city{Berlin}, \country{Germany}}}

\affil[2]{\orgdiv{Department of Electrical and Photonics Engineering}, \orgname{Technical University of Denmark}, \orgaddress{\street{2800 Kgs. Lyngby}, \country{Denmark}}}

\affil[3]{\orgdiv{Institute of Physics “Gleb Wataghin”}, \orgname{State University of Campinas}, \orgaddress{\postcode{13083-859} \city{Campinas}, \country{Brazil}}}

\abstract{\unboldmath Transition-metal dichalcogenides offer a promising platform for integrated coherent light sources, yet lasing has largely relied on hybrid photonic architectures without direct quantum-optical verification. Here, we demonstrate an all-van der Waals (all-vdW) high-$\beta$ nanobeam laser based on a WS$_2$/MoSe$_2$/WS$_{2}$ heterostructure, with the MoSe$_2$ monolayer directly integrated in the WS$_{2}$-based optical resonator for optimal gain-mode overlap. The devices exhibit efficient exciton-cavity coupling at cryogenic temperatures, strongly directional and linearly polarized emission, soft nonlinear input-output characteristics and linewidth narrowing, enabling lasing operation with $\beta$ near unity. Excitation-power-dependent photon-autocorrelation measurements reveal a transition from thermal to Poissonian photon statistics, with $g_{\mathrm{peak}}^{(2)}(0)$ decreasing from $(1.28\,\pm\,0.09)$ near threshold to $(1.07\,\pm\,0.07)$ above threshold, directly verifying lasing operation. Furthermore, temporal broadening of the autocorrelation uncovers fluctuation-dominated lasing dynamics. These results establish all-vdW heterostructures as a highly attractive platform for integrated coherent light sources in layered-material photonic architectures and scalable quantum-photonic circuits.}


\maketitle

\section{Main}\label{sec:intro}

Atomically thin semiconductors and van der Waals (vdW) heterostructures have emerged as promising platforms for integrated nanophotonics owing to their flexible assembly, ultimate thickness scaling and exceptionally strong light-matter interaction \cite{Geim2013,DeAbajo2025}. In particular, transition metal dichalcogenide (TMDC) monolayers (ML) host tightly bound excitons with large oscillator strengths and efficient optical emission at the atomic-thickness limit \cite{Mak2010}. Together with deterministic vdW stacking, these properties have enabled a wide range of photonic and optoelectronic devices, including transistors \cite{Radisavljevic2011}, photodetectors \cite{Lee2014}, electro-optic modulators \cite{Klein2019}, superconducting single-photon detectors \cite{Metuh2025superconductor} and quantum emitters \cite{Koperski2015,Srivastava2015,He2015,Paralikis2025}, establishing layered materials as a versatile platform for integrated nanophotonic technologies.

The continued miniaturization of optical cavities has simultaneously intensified the search for compact coherent light sources for optical interconnects \cite{Ma2019}, neuromorphic photonic hardware \cite{Chen2023} and quantum information processing \cite{Heindel2023}. High-$\beta$ nanolasers are particularly attractive because efficient spontaneous-emission funneling enables coherent light generation in wavelength-scale cavities approaching the diffraction limit while substantially reducing the threshold for laser operation \cite{Khajavikhan2012,Deng2021}. Their ultra-small cavity volumes further give rise to a fluctuation-dominated lasing regime in which spontaneous emission and stimulated emission coexist over an extended excitation range, fundamentally distinguishing them from conventional low-$\beta$ lasers \cite{Chow2014}.

These developments have stimulated considerable interest in combining TMDC MLs with nanophotonic resonators. Significant progress has been achieved using photonic crystal cavities \cite{Wu2015,Li2017}, distributed Bragg reflector resonators \cite{KoulasSimos2024} and microdisks \cite{Ye2015}, while electrically injected lasing has also been demonstrated recently \cite{Chen2025}. However, nearly all reported devices rely on hybrid architectures in which TMDC MLs are transferred onto pre-fabricated dielectric resonators. Consequently, optical confinement and gain remain physically separated and interact only through an evanescent optical field, limiting modal overlap, spontaneous-emission enhancement and scalable integration of active and passive photonic components.

More recently, vdW materials themselves have begun to emerge as photonic building blocks beyond their conventional role as gain media \cite{Zotev2025}. Layered TMDC heterostructures have been shown to support optical waveguiding, strongly confined cavity modes, enhanced Purcell factors and superconducting nanowire detectors \cite{Verre2019,Munkhbat2022,Binkowski2025,Metuh2025superconductor}, raising the prospect of fully integrated vdW photonic circuits \cite{Metuh2025}. However, coherent light generation has not yet been demonstrated in an all-vdW cavity architecture with direct quantum-optical verification of lasing. Previous all-vdW cavity implementations either did not reach the lasing regime \cite{Alekseev2025} or relied on indirect-gap bulk emission \cite{Sung2022}. Moreover, reported TMDC nanolasers have largely been identified through conventional power-dependent signatures, such as soft nonlinear input-output (I/O) characteristics and moderate linewidth narrowing, which cannot unambiguously distinguish high-$\beta$ lasing from amplified spontaneous emission without quantum-optical measurements of the photon statistics \cite{Strauf2006,Ulrich2007,Samuel2009,Kreinberg2017,KoulasSimos2022,KoulasSimos2024}.

Here, we demonstrate an integrated all-vdW high-$\beta$ nanolaser. The device is based on a WS$_2$/MoSe$_2$/WS$_2$ heterostructure patterned into a photonic-crystal nanobeam cavity to maximize the spatial overlap between the excitonic gain medium and the confined cavity mode while realizing a fully integrated layered-material photonic platform. Temperature-dependent microphotoluminescence (\textmu PL) spectroscopy reveals efficient exciton-cavity coupling and an approximately 70-fold intensity enhancement of the cavity-mediated emission compared with the unpatterned heterostructure, while real- and Fourier-space measurements identify the relevant nanobeam cavity mode through its strong spatial confinement and highly directional in-plane emission. Under optical excitation, the devices exhibit the characteristic classical signatures of high-$\beta$ lasing, including soft nonlinear I/O characteristics, gradual linewidth narrowing and a near-unity spontaneous-emission coupling factor. Excitation-power-dependent photon autocorrelation measurements further reveal a continuous transition from thermal to Poissonian photon statistics, with the zero-delay photon autocorrelation decreasing from $g_{\mathrm{peak}}^{(2)}(0)=(1.28\pm0.09)$ near threshold to $(1.07\pm0.07)$ above threshold, thereby providing direct quantum-optical verification of lasing. In addition, the temporal evolution of the autocorrelation functions uncovers fluctuation-dominated laser dynamics associated with delayed lasing build-up and stochastic pulse-to-pulse timing jitter in ultra-small cavities.

\section{All-van der Waals nanobeam cavity platform}\label{sec:platform}
\begin{figure}[!t]
\centering
\includegraphics[width=\textwidth]{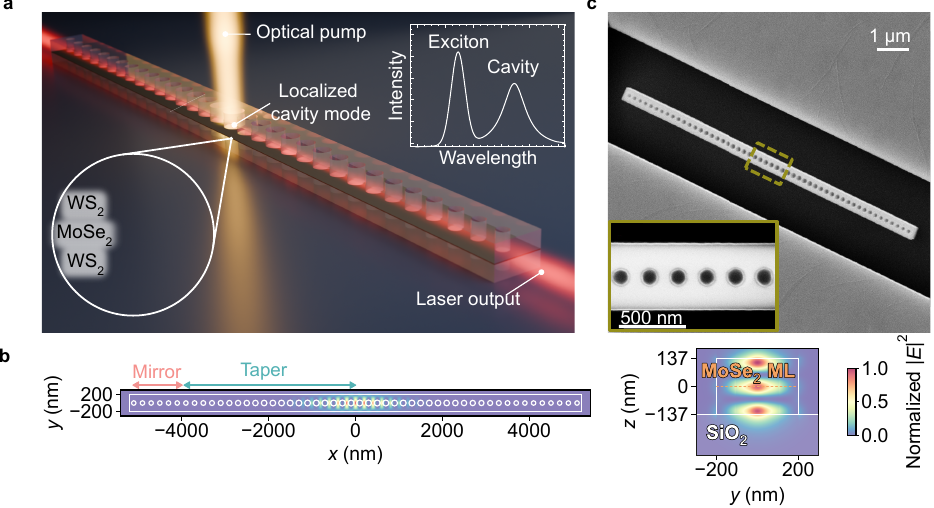}
\caption{
\textbf{All-vdW nanobeam cavity platform.} \textbf{a} Schematic illustration of the WS$_2$/MoSe$_2$/WS$_2$ heterostructure integrated into a photonic-crystal nanobeam cavity. The embedded MoSe$_2$ ML provides optical gain, while the surrounding multilayer WS$_2$ simultaneously acts as the photonic cavity. \textbf{b} Three-dimensional FEM simulations of the dielectric-band cavity mode, showing the electric-field intensity distribution in the lateral ($x$--$y$) and vertical ($y$--$z$) planes together with the nanobeam geometry consisting of a taper and a mirror section. \textbf{c} Scanning electron microscopy image of a representative fabricated nanobeam cavity with a magnified view of the photonic-crystal region.}\label{fig:Figure1}
\end{figure}

The architecture, optical design and fabrication of the all-vdW nanobeam cavity are summarized in Fig.~\ref{fig:Figure1}. The device consists of a mechanically exfoliated WS$_2$/1L-MoSe$_2$/WS$_2$ heterostructure patterned into a one-dimensional photonic-crystal nanobeam cavity (Fig.~\ref{fig:Figure1}a). The embedded MoSe$_2$ ML provides optical gain, while the surrounding WS$_2$ multilayer simultaneously ensures optical confinement and maximizes the spatial overlap between the excitonic emission and the confined cavity mode. Unlike hybrid architectures employing transferred MLs, the gain medium is directly integrated within the photonic resonator.

The optical confinement properties were investigated using three-dimensional finite-element (FEM) simulations in JCMsuite \cite{Burger2008}. The cavity was designed for spectral resonance with the A-exciton of the embedded MoSe$_2$ ML (750--770~nm) at cryogenic temperatures  \cite{Ross2013,Robert2016,Palekar2024}. The target room-temperature resonance was therefore chosen to be approximately 765~nm to account for the expected 6.5-nm blueshift upon cooling (Supplementary Section~2). The simulated electric-field distribution of the dielectric-band mode is shown in Fig.~\ref{fig:Figure1}b. The adiabatically tapered cavity along the nanobeam axis strongly localizes the optical field \cite{Quan2011}, reducing the mode volume by approximately a factor of three compared with an untapered design while confining approximately 93\% of the electromagnetic energy within the WS$_2$ nanobeam (Supplementary Section~1).

The heterostructures were assembled by deterministic dry-transfer stacking of mechanically exfoliated TMDC flakes \cite{Metuh2025}, followed by electron-beam lithography (EBL) and inductively coupled plasma reactive-ion etching (ICP-RIE) (Methods and Supplementary Section~3). A scanning electron microscopy (SEM) image of a representative fabricated nanobeam is shown in Fig.~\ref{fig:Figure1}c. Structural analysis yields beam widths and hole diameters approximately 14\% larger than the nominal design, whereas the lattice constant remains controlled within $\pm\,2.2$~nm (Supplementary Section~4). The slightly rounded hole morphology originates from the crystallographic anisotropy of the layered TMDC heterostructure during plasma etching \cite{Munkhbat2020}.

To evaluate the optical properties of the fabricated devices, the experimentally determined dimensions were incorporated into the FEM model. While the nominal geometry predicts a cavity resonance at 771~nm, the SEM-derived dimensions shift the resonance to 763~nm, in excellent agreement with the experimentally observed PL peak (Supplementary Sections~5 and~6). Although the increased beam width alone would redshift the cavity mode, the simultaneously enlarged air holes reduce the dielectric filling fraction and produce a stronger blueshift, resulting in excellent spectral overlap with the low-temperature MoSe$_2$ exciton. Using the experimentally extracted geometry, the simulations yield a quality factor of $Q=13\,670$, a mode volume of $V_{\mathrm m}=2.34\,(\lambda/n)^3$, a ML confinement factor of $\Gamma_{\mathrm{ML}}=0.80\%$ and a Purcell factor of $F_{\mathrm P}=125$. These values remain close to those of the nominal design and are superior or at least competitive with state-of-the-art hybrid photonic-crystal cavities incorporating transferred TMDC MLs (Supplementary Section~1) \cite{Wu2015,Li2017}.
\begin{figure}[!t]
\centering
\includegraphics[width=\textwidth]{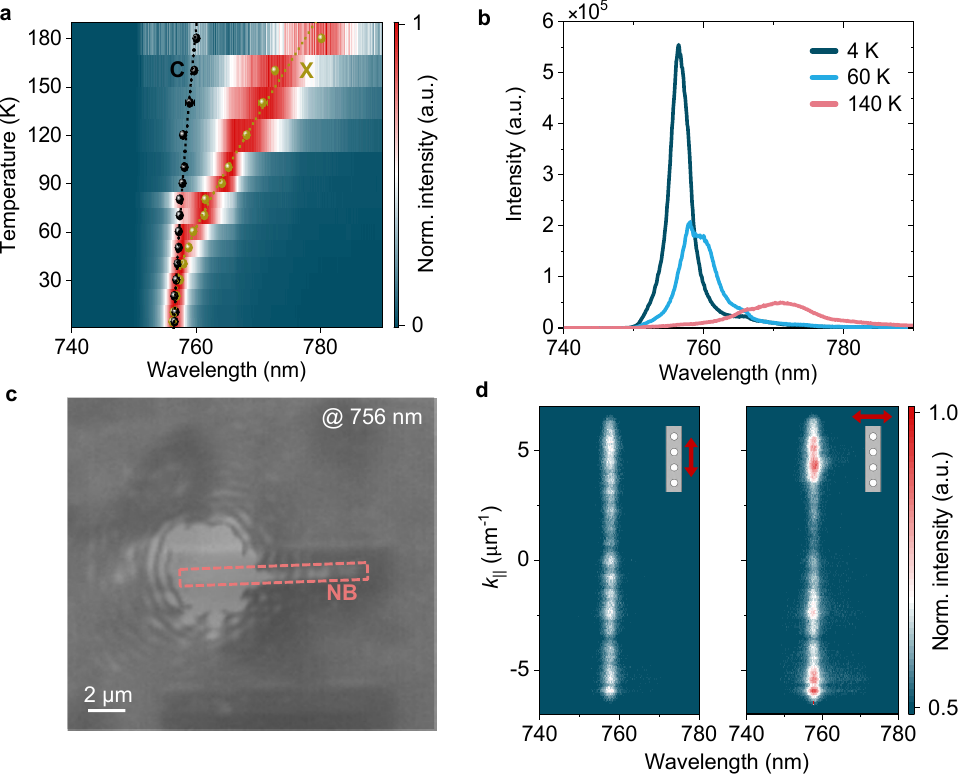}
\caption{\textbf{Exciton-cavity coupling and directional emission in all-vdW nanobeam cavities.}
\textbf{a} Normalized temperature-dependent \textmu PL map showing the spectral evolution of the neutral MoSe$_2$ A-exciton (X) and cavity mode (C). The dotted lines represent fits to the extracted emission wavelengths (see Supplementary Section 6 for details on the fits). \textbf{b} Representative unnormalized spectra recorded at selected temperatures. \textbf{c} Real-space PL image recorded at the cavity resonance wavelength, showing resonant coupling into and propagation along the nanobeam. \textbf{d} Angle-resolved Fourier-space emission measured for linear polarizations oriented perpendicular and parallel to the nanobeam axis, respectively. The pronounced high-angle emission for perpendicular polarization confirms directional radiation from the confined nanobeam cavity mode. Insets indicate the relative orientation between polarization and nanobeam axis.}\label{fig:Figure2}
\end{figure}
\section{Exciton-cavity coupling and directional emission}\label{sec:exciton-cavity}
To investigate the interaction between the embedded MoSe$_2$ ML and the confined cavity mode, temperature-dependent \textmu PL spectroscopy was performed on representative nanobeam cavities. Measurements on unpatterned WS$_2$/MoSe$_2$/WS$_2$ heterostructures establish the intrinsic optical response of the embedded ML and identify the optical transition as the neutral MoSe$_2$ A-exciton by complementary \textmu PL and \textmu refection spectroscopy (Supplementary Section~5). Relative to the unpatterned heterostructure, the nanobeam exhibits an approximately 70-fold intensity enhancement of the spectrally localized cavity emission, demonstrating efficient Purcell-induced spontaneous-emission funneling into the confined optical mode (Supplementary Section~5).

The normalized temperature-dependent spectra of a representative nanobeam cavity are shown in Fig.~\ref{fig:Figure2}a. Two resonances are observed, corresponding to the neutral MoSe$_2$ A-exciton (X) and the cavity mode (C). Fits to the extracted exciton and cavity wavelengths (solid curves in Fig.~\ref{fig:Figure2}a) quantify the temperature evolution of both resonances and the resulting exciton-cavity detuning. As the temperature increases, the exciton redshifts approximately $23.8$~nm owing to bandgap renormalization \cite{Ross2013}, whereas the cavity resonance shifts only weakly ($\approx3.5$~nm) due to the small thermo-optic response of the surrounding WS$_2$ multilayer \cite{Liu2020temperature} (Supplementary Section~6). The resulting difference in temperature coefficients enables continuous control of the exciton-cavity detuning. Figure~\ref{fig:Figure2}b shows representative unnormalized spectra at selected temperatures, illustrating the pronounced intensity enhancement near 4~K, where the exciton is resonant with the cavity mode.

The cavity-related origin of the emission is further supported by real- and Fourier-space measurements. Fig.~\ref{fig:Figure2}c shows a real-space PL image recorded at the cavity resonance wavelength, revealing light coupled into, and guided along, the nanobeam. Measurements performed away from resonance show no corresponding guided emission (Supplementary Section~7). Fourier-space spectroscopy (Fig.~\ref{fig:Figure2}d) exhibits a highly directional far-field radiation pattern with dominant high-angle emission polarized perpendicular to the nanobeam axis. The measured far-field distribution agrees closely with the FEM simulations (Supplementary Section~1), confirming mode emission at angles close to $\pm54^{\circ}$. Together, these measurements demonstrate efficient coupling between the embedded MoSe$_2$ ML and the confined nanobeam cavity mode, providing the basis for the lasing experiments discussed below.

\section{Conventional signatures of high-\texorpdfstring{$\beta$}{beta} coherent emission}\label{sec:IO-linewidth}
\begin{figure}[!t]
\centering
\includegraphics[width=\textwidth]{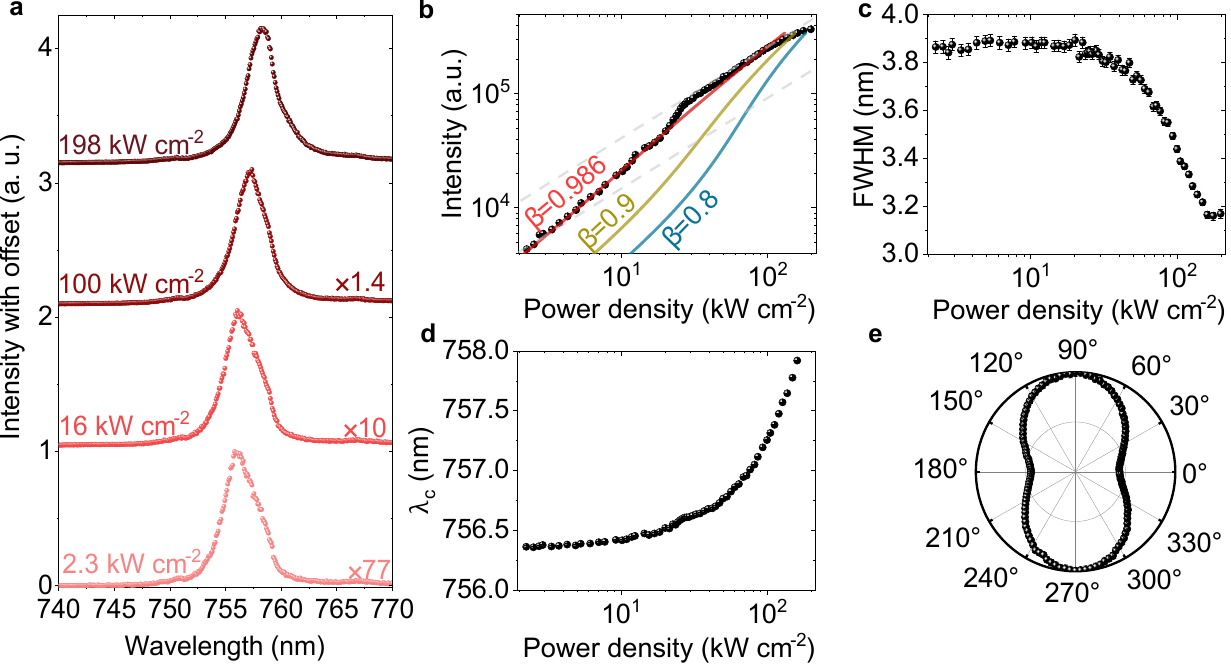}
\caption{\textbf{Demonstration of high-$\beta$ coherent emission in all-vdW nanobeam cavities.} \textbf{a} Power-dependent cavity spectra recorded under quasi-CW excitation. \textbf{b} I/O curve showing the characteristic soft nonlinear transition of a high-$\beta$ nanolaser. Solid lines show rate-equation fits for different values of $\beta$, with the best fit yielding $\beta=(0.986\,\pm\,0.007)$ and $P_{\mathrm{th}}=(9\,\pm\,1)$~kW\,cm$^{-2}$. Dashed gray lines indicate the low- and high-power linear regimes. \textbf{c} Power-dependent cavity linewidth, quantified by the full width at half maximum (FWHM) of the Voigt-profile fits, revealing moderate narrowing around the nonlinear transition. \textbf{d} Power-dependent spectral redshift of the cavity mode, attributed to local laser-induced heating. \textbf{e} Polarization-resolved cavity emission exhibiting a degree of linear polarization of $(54.7\,\pm\,0.2)\%$, consistent with the confined nanobeam cavity-mode symmetry.}\label{fig:Figure3}
\end{figure}

Having established efficient exciton-cavity coupling within the nanobeam architecture, Fig.~\ref{fig:Figure3} depicts the excitation-power-dependent emission of the embedded nanobeam cavity under quasi-continuous-wave (quasi-CW) optical excitation at 4~K. The onset of coherent emission is already evident from the evolution of the cavity spectra shown in Fig.~\ref{fig:Figure3}a. At low excitation power densities, the cavity mode appears as a broad spontaneous-emission feature. Increasing the pump power enhances the emission intensity by two orders of magnitude, while the cavity resonance simultaneously narrows and undergoes a slight redshift.

To quantify these changes, all spectra were fitted using a Voigt profile, accounting for possible lasing-induced lineshape effects \cite{KoulasSimos2022}. The extracted emission intensity, linewidth and resonance wavelength are summarized in Fig.~\ref{fig:Figure3}b-d. The I/O characteristics (Fig.~\ref{fig:Figure3}b) exhibit the characteristic soft S-shaped response indicative of a high-$\beta$ nanolaser. At low excitation power densities, the cavity emission increases approximately linearly owing to spontaneous excitonic recombination. Upon increasing the excitation density, a smooth nonlinear transition emerges before the emission enters a second nearly linear regime dominated by stimulated emission. Such gradual threshold behaviour is characteristic of high-$\beta$ nanolasers, in which a large fraction of spontaneous emission is funneled directly into the lasing mode \cite{Khajavikhan2012,Chow2018}. The measured I/O characteristics are accurately reproduced by the high-$\beta$ rate-equation model described in Supplementary Section~8 \cite{Bjork1991}. The best fit yields a spontaneous-emission coupling factor of $\beta=(0.986\,\pm\,0.007)$ and a characteristic threshold power of $P_{\mathrm{th}}=(9\,\pm\,1)$~kW\,cm$^{-2}$, indicating near-thresholdless behavior. Substantially smaller values of $\beta$ systematically fail to reproduce the measured nonlinear transition while keeping all remaining model parameters fixed, providing strong evidence that the cavity operates in the near-unity-$\beta$ regime. Owing to the extremely efficient spontaneous-emission coupling, however, the extracted threshold should be regarded as a characteristic transition power rather than a sharply defined lasing threshold.

Concurrently, the cavity linewidth remains nearly constant at low excitation powers before narrowing by approximately a factor of 1.25 around the nonlinear transition (Fig.~\ref{fig:Figure3}c), reflecting the progressive build-up of optical coherence. Unlike conventional low-$\beta$ lasers, where stimulated emission is accompanied by an abrupt linewidth collapse, the comparatively gradual narrowing observed here is expected for high-$\beta$ nanolasers, where spontaneous and stimulated emission coexist over an extended excitation range \cite{Ye2015,Wu2015,Li2017,KoulasSimos2024}. Simultaneously, the cavity resonance continuously redshifts with increasing excitation power (Fig.~\ref{fig:Figure3}d), consistent with local laser-induced heating of the nanobeam. Notably, comparison with the independently measured temperature-dependent resonance shift in a nominally identical nanobeam indicates an effective cavity temperature of approximately 60~K at the highest excitation power densities (Supplementary Section~11).

The cavity origin of the nonlinear emission is further confirmed by polarization-resolved measurements (Fig.~\ref{fig:Figure3}e). The emission exhibits a linear polarization contrast of $(54.7\,\pm\,0.2)\%$, with the dominant polarization oriented perpendicular to the nanobeam axis, in excellent agreement with both the simulated cavity-mode symmetry and the measured far-field radiation pattern (Fig.~\ref{fig:Figure2}d). This pronounced polarization anisotropy confirms that the nonlinear emission originates from the confined nanobeam cavity mode rather than from spatially extended excitonic background emission.

The robustness of the temperature- and excitation-power-dependent observations is further assessed on a second nanobeam cavity fabricated within the same WS$_2$/MoSe$_2$/WS$_2$ heterostructure (Supplementary Section~11). The key high-$\beta$ signatures are excellently reproduced in the second device, yielding $\beta=(0.987\,\pm\,0.004)$ and $P_{\mathrm{th}}=(29\,\pm\,1)$~kW\,cm$^{-2}$. This confirms that the observed high-$\beta$ behaviour is reproducible, rather than device-specific.

Taken together, the soft nonlinear I/O characteristics, near-unity spontaneous-emission coupling factor, gradual linewidth narrowing, excitation-induced cavity redshift and pronounced linear polarization provide compelling evidence for coherent emission from the all-vdW nanobeam cavities. Still, these conventional signatures alone cannot unambiguously distinguish high-$\beta$ lasing from amplified spontaneous emission in fluctuation-dominated nanocavities \cite{Strauf2006,Ulrich2007,Samuel2009,Kreinberg2017,KoulasSimos2022,KoulasSimos2024}. Direct quantum-optical verification therefore requires photon-correlation measurements, which are presented for the relevant representative device in the following section.

\section{Quantum optical verification of high-\texorpdfstring{$\beta$}{beta} lasing and fluctuation-driven lasing dynamics in photon-autocorrelation measurements}\label{sec:g2}
\begin{figure}[!t]
\centering
\includegraphics[width=\textwidth]{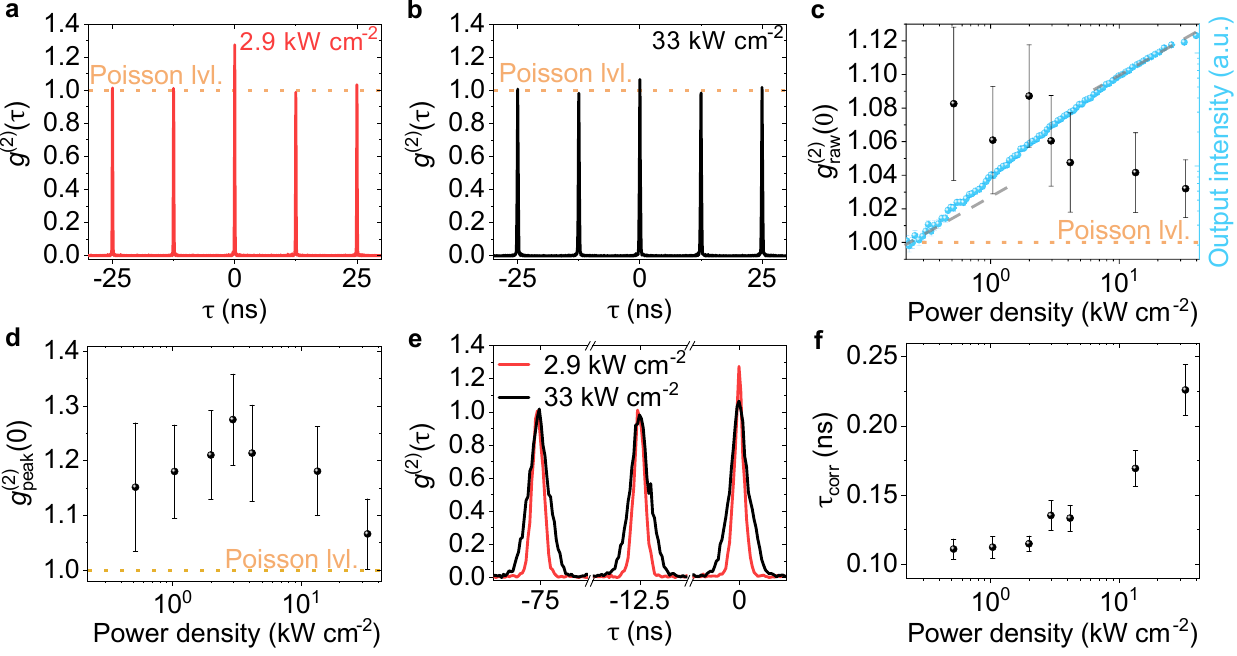}
\caption{\textbf{Quantum-optical verification of high-$\beta$ lasing in embedded all-vdW nanobeam cavities.} \textbf{a, b} Representative second-order autocorrelation histograms $g^{(2)}(\tau)$ measured close to laser threshold and far above threshold under pulsed optical excitation, revealing the transition from thermal to Poissonian photon statistics. \textbf{c} Excitation-power-density-dependent evolution of the integrated autocorrelation value $g_{\mathrm{raw}}^{(2)}(0)$ extracted from the ratio between the area of the zero-delay peak and the mean area of the adjacent side peaks, shown together with the pulsed I/O characteristics. The horizontal dashed line indicates the Poisson limit. Dashed gray lines indicate the low- and high-power linear regimes. \textbf{d} Corresponding evolution of the normalized zero-delay peak amplitudes $g_{\mathrm{peak}}^{(2)}(0)$ as a function of excitation power density. \textbf{e} Representative autocorrelation peaks recorded at selected excitation power densities, illustrating the excitation-induced temporal broadening of the correlation function above the nonlinear transition. \textbf{f} Temporal FWHM $\tau_{\mathrm{corr}}$ of the autocorrelation peaks as a function of excitation power, revealing systematic broadening of the correlation dynamics above threshold.}\label{fig:Figure4}
\end{figure}
To unambiguously verify high-$\beta$ lasing in the investigated nanobeam device, we probe the photon statistics of the emitted light by performing excitation-power-dependent second-order photon-autocorrelation measurements using a Hanbury Brown and Twiss interferometer equipped with superconducting nanowire single-photon detectors. In contrast to the quasi-CW excitation employed for the spectroscopic characterization, these measurements were performed under well-defined pulsed excitation, which provides a robust determination of the zero-delay second-order correlation function by temporally separating successive excitation events. Representative autocorrelation histograms recorded near and well above the lasing threshold are shown in Fig.~\ref{fig:Figure4}a and b, respectively. At low excitation power densities, the enhanced zero-delay peak reveals photon bunching characteristic of partially thermal emission dominated by spontaneous-emission fluctuations \cite{Strauf2006,Ulrich2007,KoulasSimos2024}. With increasing excitation power, the bunching gradually disappears and the correlation histogram approaches the Poissonian limit expected for coherent emission. To quantify this evolution, we first extract the zero-delay autocorrelation from the ratio between the integrated area of the central peak and the average integrated area of the adjacent side peaks. The resulting excitation-power dependence of $g_{\mathrm{raw}}^{(2)}(0)$ is shown together with the corresponding pulsed I/O characteristics in Fig.~\ref{fig:Figure4}c. Across the nonlinear transition, $g_{\mathrm{raw}}^{(2)}(0)$ continuously decreases from $(1.09\,\pm\,0.03)$ to $(1.03\,\pm\,0.02)$, demonstrating the progressive suppression of spontaneous-emission fluctuations as stimulated emission becomes dominant. Because the measurement resolution (37~ps) is below the emission coherence time, the observed bunching represents a lower bound of the intrinsic photon bunching \cite{Kim2016}. Nevertheless, the continuous evolution towards the Poissonian limit provides direct quantum-optical verification of coherent light generation in the all-vdW nanobeam cavity. As a complementary analysis, the normalized height of the zero-delay correlation peak is also evaluated (Fig.~\ref{fig:Figure4}d). This independent method yields the same qualitative behavior, with $g_{\mathrm{peak}}^{(2)}(0)$ decreasing from $(1.28\,\pm\,0.09)$ near the threshold to $(1.07\,\pm\,0.07)$ deep within the stimulated-emission regime. Although the peak-height analysis is inherently more sensitive to the temporal profile of the correlation peaks than the integrated-area approach, the excellent agreement between both methods confirms the continuous transition from thermal to coherent photon statistics.

Beyond the photon statistics, the temporal structure of the autocorrelation histograms provides additional insight into the lasing dynamics. Representative autocorrelation peaks recorded below and above the threshold are compared in Fig.~\ref{fig:Figure4}e. Above threshold, the peaks become noticeably broader, indicating that the temporal characteristics of the emitted pulses evolve together with the photon statistics. The extracted full width at half maximum $\tau_\mathrm{corr}$ is illustrated in Fig.~\ref{fig:Figure4}f, revealing a systematic increase from $(0.11\,\pm\,0.01)$~ns below threshold to $(0.23\,\pm\,0.02)$~ns at the highest excitation power densities. Such excitation-power-dependent broadening is qualitatively consistent with delayed lasing build-up and dynamical hysteresis predicted and observed in pulsed high-$\beta$ nanolasers \cite{Pan2016}. An alternative explanation based on changes in the radiative lifetime can be excluded. Independent streak-camera measurements reveal a resolution-limited emission response below 6~ps throughout the investigated excitation range (Supplementary Section~9), demonstrating that the observed broadening occurs on timescales more than an order of magnitude larger than the intrinsic radiative dynamics and therefore cannot originate from long-lived excitonic or defect-assisted recombination. A more detailed statistical analysis presented in Supplementary Section~10 further reveals that the non-zero-delay peaks become systematically broader than the central peak in the stimulated-emission regime. This behavior is consistent with spontaneous-emission-driven pulse-to-pulse timing jitter of the lasing turn-on \cite{Lebreton2015}. While photon pairs contributing to the central peak experience a single timing fluctuation, correlations between different excitation cycles accumulate timing uncertainty from two statistically independent lasing events, naturally leading to broader side peaks.

Taken together, the continuous evolution from thermal towards Poissonian photon statistics, together with the characteristic temporal broadening and pulse-to-pulse timing jitter observed in the autocorrelation measurements, provides direct quantum-optical verification of coherent light generation in the ML-embedded all-vdW nanobeam cavity. Beyond confirming stimulated emission, these measurements demonstrate that the devices operate deeply within the fluctuation-dominated high-$\beta$ regime, where spontaneous-emission fluctuations continue to influence the laser dynamics well above the nonlinear transition.

\section{Conclusions} \label{sec:conclusions}
We have demonstrated an all-vdW high-$\beta$ nanobeam cavity laser based on a WS$_2$/1L-MoSe$_2$/WS$_2$ heterostructure. In contrast to previous hybrid implementations, the excitonic gain medium is embedded within the optical resonator, maximizing the overlap between the confined cavity mode and the active ML while realizing a fully integrated layered-material photonic platform. With a $\beta$ factor of $(0.986\,\pm\,0.007)$ the devices reach the limit of thresholdless lasing where excitation-power-dependent photon-correlation measurements provide direct quantum-optical verification of the transition from thermal to coherent emission. Beyond establishing coherent light generation, the observed autocorrelation dynamics reveal that the devices operate deeply within the fluctuation-dominated high-$\beta$ regime, where delayed lasing build-up and spontaneous-emission-driven pulse-to-pulse timing jitter remain relevant well above the nonlinear transition. These results establish quantum-optically verified lasing in an all-vdW nanolaser employing an embedded ML gain medium. They further demonstrate layered TMDC heterostructures can simultaneously provide optical gain, photonic confinement and coherent light generation within a single material system, establishing them as a viable platform for integrated vdW nanophotonics. The present proof-of-principle devices were realized using mechanically exfoliated materials and deterministic transfer assembly. Future developments based on wafer-scale TMDC heterostructures, improved optical quality through hexagonal boron nitride encapsulation and electrically injected operation should enhance device performance while enabling scalability. Combined with recently demonstrated vdW waveguides, resonators, quantum emitters and superconducting nanowire single-photon detectors, the demonstrated nanobeam laser provides a key building block towards fully integrated layered-material quantum-photonic circuits.

\backmatter
\section*{Methods}
\addcontentsline{toc}{section}{Methods}

\subsection*{Numerical simulations}
\addcontentsline{toc}{subsection}{Numerical simulations}
The optical properties of the nanobeam cavities were investigated using three-dimensional FEM simulations. The cavity geometry was constructed using the structural dimensions extracted from SEM images of the fabricated devices, including the nanobeam width, lattice constant and hole dimensions. The optical response of the WS$_2$/MoSe$_2$/WS$_2$ heterostructure was modeled using literature values for the refractive indices of the constituent materials at room temperature \cite{Munkhbat2022optical}. The eigenmodes of the photonic crystal cavity were calculated to determine the resonance wavelength, quality factor and spatial field distribution of the fundamental cavity mode. The mode volume was evaluated from the simulated electric-field distribution according to $V_\mathrm{m}=\frac{\int \varepsilon(\mathbf{r})|E(\mathbf{r})|^2\,dV}{\max[\varepsilon(\mathbf{r})|E(\mathbf{r})|^2]}$, where $\varepsilon(\mathbf{r})$ denotes the local dielectric permittivity and $E(\mathbf{r})$ the electric field. The Purcell factor was calculated by employing a resonance expansion algorithm \cite{Binkowski2025}. The overlap between the cavity mode and the embedded MoSe$_2$ ML was evaluated from the simulated electric-field distribution within the active region. Far-field radiation patterns were obtained by projecting the simulated cavity fields into momentum space and were compared with the experimentally measured Fourier-space emission profiles.

\subsection*{Device fabrication}
\addcontentsline{toc}{subsection}{Device fabrication}
The WS$_2$/MoSe$_2$/WS$_2$ heterostructures were assembled using a deterministic dry-transfer technique. WS$_2$ multilayer flakes and MoSe$_2$ ML were mechanically exfoliated from bulk crystals (HQ Graphene) onto polydimethylsiloxane (PDMS) stamps. Suitable WS$_2$ flakes were selected by measuring their thickness via spectral reflectivity measurements from confocal illumination with a broadband source and subsequent fitting via the transfer matrix method, in order to obtain a total heterostructure thickness of approximately 275~nm. MoSe$_2$ ML flakes were identified from their optical contrast under optical microscopy, which was verified by observing their room-temperature PL with a blue LED. 
To fabricate the heterostructure, a bottom WS$_2$ multilayer flake was first transferred onto a Si substrate covered by a 2 \textmu m-thick thermally grown SiO$_2$ layer. Subsequently, a MoSe$_2$ ML and a second WS$_2$ multilayer flake were sequentially transferred to form the vertically embedded WS$_2$/MoSe$_2$/WS$_2$ heterostructure. Each flake was aligned and transferred using a transfer station (HQ Graphene Systems HQ2D MAN). Following the heterostructure transfer, the sample was cleaned using acetone, isopropanol and deionized water to remove residual contaminants. The fabrication procedure follows our previously reported approach for vdW photonic devices \cite{Metuh2025}. Nanobeam cavities were fabricated directly from the assembled heterostructures using EBL and ICP-RIE. The heterostructure was coated with a positive-tone electron-beam resist (AR-P 6200, Allresist) and patterned by EBL (JBX-9500FS, JEOL, at 100~kV). After development (ZED-N50, Zeon), descumming in oxygen plasma, and a soft bake step, the pattern was transferred into the heterostructure using ICP-RIE with a CHF$_3$/Ar/CF$_4$/O$_2$ plasma chemistry (10 mTorr chamber pressure, 125~W ICP power and 25~W RF power). Finally, the residual resist mask was removed (AR 600-71, Allresist), leaving free-standing nanobeam cavity structures defined entirely within the vdW heterostructure. The fabricated devices were inspected using optical microscopy to verify the structural integrity and pattern fidelity of the nanobeam cavities prior to optical characterization. To avoid charging effects and other structural changes with a strong exposure, SEM was performed on similar samples without optically active layers.

\subsection*{Experimental configuration}
\addcontentsline{toc}{subsection}{Experimental configuration}
The experimental configuration used for the optical measurements consists of a confocal \textmu PL setup equipped with spectral, momentum-space, time-resolved and photon-correlation detection paths. The sample is mounted in a closed-cycle cryostat, where the temperature is controlled using a PID temperature controller. Optical excitation is provided either by a quasi-CW laser for all \textmu PL-measurements or by a pulsed laser source for time-resolved and photon-correlation experiments. The excitation laser is focused onto the sample using a low-temperature microscope objective with a numerical aperture of $\mathrm{NA}=0.81$, a focal length of $f=2.89$~mm and a clear aperture of approximately 4.7~mm. The PL is collected confocally through the same objective. For spectral characterization, the collected emission is directed to a monochromator and detected using a charge-coupled device camera. Excitation-power-dependent spectra are recorded by varying the incident laser power using calibrated neutral-density filters. Polarization-resolved measurements are performed by placing a rotatable $\lambda /2$-plate and linear polarizer in the collection path. Temperature-dependent measurements are carried out by stabilizing the cryostat at each temperature before spectral acquisition. Angle-resolved measurements are performed using a Fourier-imaging configuration that maps the back focal plane of the objective onto the entrance plane of the monochromator. Owing to the high numerical aperture of the objective, emission angles up to $\theta_{\mathrm{max}}=\arcsin(0.81)\approx54.1^\circ$ can be collected. The corresponding accessible in-plane wavevector range is given by $k_{\parallel,\mathrm{max}}=\frac{2\pi}{\lambda}\mathrm{NA}$, which corresponds to $k_{\parallel,\mathrm{max}}\approx6.7~\mu\mathrm{m}^{-1}$ at $\lambda\approx760$~nm. The objective clear aperture is consistent with the expected back-focal-plane diameter $D=2f\mathrm{NA}\approx4.7$~mm. 

For photon-autocorrelation measurements, the emission is spectrally filtered using a bandpass filter with a transmission window of approximately ($\pm5$)~nm around the investigated cavity mode and then coupled into a fiber-based Hanbury Brown and Twiss setup. The signal is split into two fiber arms and detected by two superconducting nanowire single-photon detectors. The photon arrival times are recorded using time-correlated single-photon counting electronics to obtain excitation-power-dependent second-order autocorrelation histograms $g^{(2)}(\tau)$. The HBT setup has a combined resolution of 37~ps. In contrast to the previous configuration employing the monochromator in zero-order reflection \cite{KoulasSimos2024}, the present measurements use direct spectral filtering before fiber coupling to isolate the cavity mode while avoiding excessively narrow spectral filtering. Time-resolved PL measurements are performed using an additional streak-camera detection path. The streak camera is operated with a synchroscan unit to enable synchronized measurements under pulsed excitation. The temporal resolution of the streak-camera configuration is approximately 6~ps, allowing us to verify that the emission response of the nanobeam cavities is resolution-limited and to exclude long-lived radiative decay channels on the timescale of the photon-correlation broadening.

\bmhead{Supplementary information} Supplementary Information provides additional experimental, numerical and theoretical details supporting the results presented in the main text, including cavity design and numerical simulations, device fabrication and characterization, temperature-dependent optical spectroscopy and thermo-optic analysis, laser rate-equation modelling, complementary streak-camera and photon-correlation measurements, and reproducibility studies.

\bmhead{Acknowledgements and Funding} A. K.-S., M. K. and S.R. acknowledge funding by the Senate of Berlin, within the Program for the Promotion of Research, Innovation and Technology (ProFIT) cofinanced by the European Regional Development Fund (ERDF, Application No. 0206824, SQALE). I. M. acknowledges funding by the Senate of Berlin via Berlin Quantum. S. R. acknowledges funding from the German Research Foundation via grant RE2974/42-1 (project ID: 576694304). P.M., A.P., and B.M. acknowledge support from the European Research Council (ERC-StG “TuneTMD”, grant no. 101076437) and Villum Fonden (project no. VIL53033). The authors also acknowledge the cleanroom facilities at the Danish National Centre for Nano Fabrication and Characterization (DTU Nanolab). The authors further acknowledge support for the numerical simulations by Dr. Sven Burger and Dr. Felix Binkowski of Zuse Institute Berlin.

\bmhead{Data availability} The data supporting the findings of this study are available within the main manuscript and its supporting information files. Additional data are available from the corresponding authors upon reasonable request.

\bmhead{Conflict of interest/Competing interests}
The authors declare they have no competing interests.

\bmhead{Author contribution}
S.R., A.K.-S. and B.M. conceived the project. A.K.-S. designed the nanobeam cavities and performed the numerical simulations. P.M. and A.P. assembled the heterostructures and fabricated the devices.  P.M. carried out the scanning electron microscopy characterization. A.K.-S. performed the optical spectroscopy with support from K.G., M.K., I.L., B.L.T.R. and C.C.P. A.K.-S. performed the photon-correlation and streak-camera measurements. A.K.-S. analyzed the data with input from all authors. A.K.-S. wrote the original manuscript. S.R. and B.M. supervised the project and acquired funding. All authors discussed the results, contributed to the interpretation of the data, revised the manuscript and approved the final version.

\bibliography{References}

@article{Geim2013,
  doi = {10.1038/nature12385},
  year = {2013},
  month = jul,
  publisher = {Springer Science and Business Media {LLC}},
  volume = {499},
  number = {7459},
  pages = {419–425},
  author = {A. K. Geim and I. V. Grigorieva},
  title = {Van der {Waals} heterostructures},
  journal = {Nature}
}

@article{DeAbajo2025,
  title={Roadmap for photonics with {2D} materials},
  author={{Garc{\'\i}a de Abajo}, F Javier and Basov, Dmitri N and Koppens, Frank HL and Orsini, Lorenzo and Ceccanti, Matteo and Castilla, Sebasti{\'a}n and Cavicchi, Lorenzo and Polini, Marco and Gon{\c{c}}alves, Paulo Andr{\'e} D and Costa, Antonio T and others},
  journal={ACS Photonics},
  volume={12},
  number={8},
  pages={3961--4095},
  year={2025},
  publisher={ACS Publications},
  doi = {10.1021/acsphotonics.5c00353}
}

@article{Mak2010,
  doi = {10.1103/physrevlett.105.136805},
  year = {2010},
  month = sep,
  publisher = {American Physical Society ({APS})},
  volume = {105},
  number = {13},
  pages= {136805},
  author = {Kin Fai Mak and Changgu Lee and James Hone and Jie Shan and Tony F. Heinz},
  title = {Atomically Thin {MoS$_2$}: A New Direct-Gap Semiconductor},
  journal = {Physical Review Letters}
}

@article{Radisavljevic2011,
  doi = {10.1038/nnano.2010.279},
  year = {2011},
  month = jan,
  publisher = {Springer Science and Business Media {LLC}},
  volume = {6},
  number = {3},
  pages = {147–150},
  author = {B. Radisavljevic and A. Radenovic and J. Brivio and V. Giacometti and A. Kis},
  title = {Single-layer {MoS}$_2$ transistors},
  journal = {Nature Nanotechnology}
}

@article{Lee2014,
  title={Atomically thin p--n junctions with van der {Waals} heterointerfaces},
  author={Lee, Chul-Ho and Lee, Gwan-Hyoung and Van Der Zande, Arend M and Chen, Wenchao and Li, Yilei and Han, Minyong and Cui, Xu and Arefe, Ghidewon and Nuckolls, Colin and Heinz, Tony F and Guo, Jing and Hone, James and Kim, Philip},
  journal={Nature Nanotechnology},
  volume={9},
  number={9},
  pages={676--681},
  year={2014},
  publisher={Nature Publishing Group UK London},
  doi = {10.1038/nnano.2014.150}
}

@article{Klein2019,
  title={{2D} semiconductor nonlinear plasmonic modulators},
  author={Klein, Matthew and Badada, Bekele H and Binder, Rolf and Alfrey, Adam and McKie, Max and Koehler, Michael R and Mandrus, David G and Taniguchi, Takashi and Watanabe, Kenji and LeRoy, Brian J and Schaibley, John R.},
  journal={Nature Communications},
  volume={10},
  number={1},
  pages={3264},
  year={2019},
  publisher={Nature Publishing Group UK London},
  doi = {10.1038/s41467-019-11186-w}
}

@article{Metuh2025superconductor,
  title={Toward single-photon detection with superconducting niobium diselenide nanowires},
  author={Metuh, Pietro and Paralikis, Athanasios and Wyborski, Pawe{\l} and Jamo, Sherwan and Palermo, Alessandro and Zugliani, Lucio and Barbone, Matteo and M\"uller, Kai and Gregersen, Niels and Vaitiekenas, Saulius and Finley, Jonathan J. and Munkhbat Battulga},
  journal={ACS Photonics},
  volume={12},
  number={11},
  pages={5912--5920},
  year={2025},
  publisher={ACS Publications},
  doi= {10.1021/acsphotonics.5c01195}
}

@article{Koperski2015,
  title={Single photon emitters in exfoliated {WSe$_2$} structures},
  author={Koperski, Maciej and Nogajewski, K and Arora, Ashish and Cherkez, V and Mallet, Paul and Veuillen, J-Y and Marcus, J and Kossacki, Piotr and Potemski, M},
  journal={Nature Nanotechnology},
  volume={10},
  number={6},
  pages={503--506},
  year={2015},
  publisher={Nature Publishing Group UK London},
  doi={10.1038/nnano.2015.67}
}

@article{Srivastava2015,
  title={Optically active quantum dots in monolayer {WSe$_2$}},
  author={Srivastava, Ajit and Sidler, Meinrad and Allain, Adrien V and Lembke, Dominik S and Kis, Andras and {\.I}mamo{\u{g}}lu, Ata{\c{c}}},
  journal={Nature Nanotechnology},
  volume={10},
  number={6},
  pages={491--496},
  year={2015},
  publisher={Nature Publishing Group UK London},
  doi={10.1038/nnano.2015.60}
}

@article{He2015,
  title={Single quantum emitters in monolayer semiconductors},
  author={He, Yu-Ming and Clark, Genevieve and Schaibley, John R and He, Yu and Chen, Ming-Cheng and Wei, Yu-Jia and Ding, Xing and Zhang, Qiang and Yao, Wang and Xu, Xiaodong and Lu, Chao-Yang and Pan, Jian-Wei},
  journal={Nature Nanotechnology},
  volume={10},
  number={6},
  pages={497--502},
  year={2015},
  publisher={Nature Publishing Group UK London},
  doi={10.1038/nnano.2015.75}
}

@article{Paralikis2025,
  title={Tunable and low-noise {WSe$_2$} quantum emitters for quantum photonics},
  author={Paralikis, Athanasios and Wyborski, Pawe{\l} and Metuh, Pietro and Gregersen, Niels and Munkhbat, Battulga},
  journal={PRX Quantum},
  volume={6},
  number={4},
  pages={040339},
  year={2025},
  publisher={APS},
  doi={10.1103/cynh-ql3j}
}

@article{Ma2019,
  title={Applications of nanolasers},
  author={Ma, Ren-Min and Oulton, Rupert F},
  journal={Nature Nanotechnology},
  volume={14},
  number={1},
  pages={12--22},
  year={2019},
  publisher={Nature Publishing Group UK London},
  doi={10.1038/s41565-018-0320-y}
}

@article{Chen2023,
  title={Deep learning with coherent VCSEL neural networks},
  author={Chen, Zaijun and Sludds, Alexander and Davis III, Ronald and Christen, Ian and Bernstein, Liane and Ateshian, Lamia and Heuser, Tobias and Heermeier, Niels and Lott, James A and Reitzenstein, Stephan and others},
  journal={Nature Photonics},
  volume={17},
  number={8},
  pages={723--730},
  year={2023},
  publisher={Nature Publishing Group UK London},
  doi = {10.1038/s41566-023-01233-w}
}

@article{Heindel2023,
author = {Tobias Heindel and Je-Hyung Kim and Niels Gregersen and Armando Rastelli and Stephan Reitzenstein},
journal = {Advances in Optics and Photonics},
number = {3},
pages = {613--738},
publisher = {Optica Publishing Group},
title = {Quantum dots for photonic quantum information technology},
volume = {15},
month = {Sep},
year = {2023},
doi = {10.1364/AOP.490091},
}

@Article{Khajavikhan2012,
  author    = {Khajavikhan, M and Simic, A and Katz, M and Lee, JH and Slutsky, B and Mizrahi, A and Lomakin, V and Fainman, Y},
  journal   = {Nature},
  title     = {Thresholdless nanoscale coaxial lasers},
  year      = {2012},
  number    = {7384},
  pages     = {204–207},
  volume    = {482},
  publisher = {Nature Publishing Group},
  doi= {10.1038/nature10840}
}

@Article{Deng2021,
  author    = {Deng, Hui and Lippi, Gian Luca and M{\o}rk, Jesper and Wiersig, Jan and Reitzenstein, Stephan},
  journal   = {Advanced Optical Materials},
  title     = {Physics and Applications of High-$\beta$ Micro-and Nanolasers},
  year      = {2021},
  number    = {19},
  pages     = {2100415},
  volume    = {9},
  publisher = {Wiley Online Library},
  doi= {10.1002/adom.202100415}
}

@Article{Chow2014,
  author    = {Chow, Weng W and Jahnke, Frank and Gies, Christopher},
  journal   = {Light: Science \& Applications},
  title     = {Emission properties of nanolasers during the transition to lasing},
  year      = {2014},
  number    = {8},
  pages     = {{e201}},
  volume    = {3},
  doi       = {10.1038/lsa.2014.82},
  publisher = {Nature Publishing Group},
}

@article{Wu2015,
  doi = {10.1038/nature14290},
  year = {2015},
  month = mar,
  publisher = {Springer Science and Business Media {LLC}},
  volume = {520},
  number = {7545},
  pages = {69–72},
  author = {Sanfeng Wu and Sonia Buckley and John R. Schaibley and Liefeng Feng and Jiaqiang Yan and David G. Mandrus and Fariba Hatami and Wang Yao and Jelena Vu{\v{c}}kovi{\'{c}} and Arka Majumdar and Xiaodong Xu},
  title = {Monolayer semiconductor nanocavity lasers with ultralow thresholds},
  journal = {Nature}
}

@article{Li2017,
  doi = {10.1038/nnano.2017.128},
  year = {2017},
  month = jul,
  publisher = {Springer Science and Business Media {LLC}},
  volume = {12},
  number = {10},
  pages = {987–992},
  author = {Yongzhuo Li and Jianxing Zhang and Dandan Huang and Hao Sun and Fan Fan and Jiabin Feng and Zhen Wang and C.-Z. Ning},
  title = {Room-temperature continuous-wave lasing from monolayer molybdenum ditelluride integrated with a silicon nanobeam cavity},
  journal = {Nature Nanotechnology}
}

@article{KoulasSimos2024,
  title={High-$\beta$ lasing in self-assembled photonic-defect microcavities with a transition metal dichalcogenide monolayer as active material},
  author={Koulas-Simos, Aris and Palekar, Chirag C and Gaur, Kartik and Limame, Imad and Shih, Ching-Wen and Rosa, B{\'a}rbara L.T. and Ning, Cun-Zheng and Reitzenstein, Stephan},
  journal={Laser \& Photonics Reviews},
  volume={18},
  number={11},
  pages={2400271},
  year={2024},
  publisher={Wiley Online Library},
  doi={10.1002/lpor.202400271}
}

@article{Ye2015,
  doi = {10.1038/nphoton.2015.197},
  year = {2015},
  month = oct,
  publisher = {Springer Science and Business Media {LLC}},
  volume = {9},
  number = {11},
  pages = {733–737},
  author = {Yu Ye and Zi Jing Wong and Xiufang Lu and Xingjie Ni and Hanyu Zhu and Xianhui Chen and Yuan Wang and Xiang Zhang},
  title = {Monolayer excitonic laser},
  journal = {Nature Photonics}
}

@article{Chen2025,
  title={Electrically-Driven {2D} Semiconductor Microcavity Laser},
  author={Chen, Zheng-Zhe and Lin, Hsiang-Ting and Chang, Chiao-Yun and Adil, Muhammad and Tsai, Po-Cheng and Kao, Tsung Sheng and Chen, Chi and Lin, Shih-Yen and Shih, Min-Hsiung},
  journal={Advanced Materials},
  volume={37},
  number={42},
  pages={e09861},
  year={2025},
  publisher={Wiley Online Library},
  doi={10.1002/adma.202509861}
}

@article{Zotev2025,
  title={Nanophotonics with multilayer van der {Waals} materials},
  author={Zotev, Panaiot G and Bouteyre, Paul and Wang, Yadong and Randerson, Sam A and Hu, Xuerong and Sortino, Luca and Wang, Yue and Shegai, Timur and Gong, Su-Hyun and Tittl, Andreas and Aharonovich, Igor and Tartakovskii, Alexander I.},
  journal={Nature Photonics},
  volume={19},
  number={8},
  pages={788--802},
  year={2025},
  publisher={Nature Publishing Group UK London},
  doi={10.1038/s41566-025-01717-x}
}

@article{Verre2019,
  title={Transition metal dichalcogenide nanodisks as high-index dielectric {Mie} nanoresonators},
  author={Verre, Ruggero and Baranov, Denis G and Munkhbat, Battulga and Cuadra, Jorge and K{\"a}ll, Mikael and Shegai, Timur},
  journal={Nature Nanotechnology},
  volume={14},
  number={7},
  pages={679--683},
  year={2019},
  publisher={Nature Publishing Group UK London},
  doi={10.1038/s41565-019-0442-x}
}

@article{Munkhbat2022,
  doi = {10.1002/lpor.202200057},
  year = {2022},
  month = nov,
  publisher = {Wiley},
  volume = {17},
  number = {1},
  pages= {2200057},
  author = {Battulga Munkhbat and Bet\"{u}l K\"{u}{\c{c}}\"{u}k\"{o}z and Denis G. Baranov and Tomasz J. Antosiewicz and Timur O. Shegai},
  title = {Nanostructured Transition Metal Dichalcogenide Multilayers for Advanced Nanophotonics},
  journal = {Laser {\&} Photonics Reviews}
}

@article{Binkowski2025,
  title={High Purcell enhancement in {all-TMDC} nanobeam resonator designs with active monolayers for nanolasers},
  author={Binkowski, Felix and Koulas-Simos, Aris and Betz, Fridtjof and Plock, Matthias and Sekulic, Ivan and Manley, Phillip and Hammerschmidt, Martin and Schneider, Philipp-Immanuel and Zschiedrich, Lin and Munkhbat, Battulga and Reitzenstein, Stephan and Burger, Sven},
  journal={Physical Review B},
  volume={112},
  number={23},
  pages={235410},
  year={2025},
  publisher={APS},
  doi={10.1103/nxh9-dhvx}
}

@article{Metuh2025,
  title={Integrated on-chip quantum light sources on a van der {Waals} platform},
  author={Metuh, Pietro and Wyborski, Pawe{\'L} and Paralikis, Athanasios and Stenger, Nicolas and Gregersen, Niels and Munkhbat, Battulga},
  journal={arXiv preprint arXiv:2512.15337},
  doi={10.48550/arXiv.2512.15337},
  year={2025}
}

@article{Alekseev2025,
  title={Engineering whispering gallery modes in {MoSe$_2$/WS$_2$} double heterostructure nanocavities: Towards developing all-{TMDC} light sources},
  author={Alekseev, PA and Milekhin, IA and Gasnikova, KA and Eliseyev, IA and Davydov, V Yu and Bogdanov, AA and Kravtsov, Vasily and Mikhin, AO and Borodin, BR and Milekhin, AG},
  journal={Materials Today Nano},
  volume={30},
  pages={100633},
  year={2025},
  publisher={Elsevier},
  doi={10.1016/j.mtnano.2025.100633}
}

@article{Sung2022,
  title={Room-temperature continuous-wave indirect-bandgap transition lasing in an ultra-thin {WS$_2$} disk},
  author={Sung, Junghyun and Shin, Dongjin and Cho, HyunHee and Lee, Seong Won and Park, Seungmin and Kim, Young Duck and Moon, Jong Sung and Kim, Je-Hyung and Gong, Su-Hyun},
  journal={Nature Photonics},
  volume={16},
  number={11},
  pages={792--797},
  year={2022},
  publisher={Nature Publishing Group UK London},
  doi={10.1038/s41566-022-01085-w}
}

@Article{Strauf2006,
  author    = {Strauf, S and Hennessy, K and Rakher, MT and Choi, Y-S and Badolato, A and Andreani, LC and Hu, EL and Petroff, PM and Bouwmeester, D},
  journal   = {Physical Review Letters},
  title     = {Self-tuned quantum dot gain in photonic crystal lasers},
  year      = {2006},
  number    = {12},
  pages     = {127404},
  volume    = {96},
  publisher = {APS},
  doi={10.1103/PhysRevLett.96.127404}
}

@Article{Ulrich2007,
  author    = {Ulrich, SM and Gies, C and Ates, S and Wiersig, J and Reitzenstein, S and Hofmann, C and L{\"o}ffler, A and Forchel, A and Jahnke, F and Michler, P},
  journal   = {Physical Review Letters},
  title     = {Photon statistics of semiconductor microcavity lasers},
  year      = {2007},
  number    = {4},
  pages     = {043906},
  volume    = {98},
  publisher = {APS},
  doi = {10.1103/PhysRevLett.98.043906}
}

@Article{Samuel2009,
  author    = {Samuel, Ifor DW and Namdas, Ebinazar B and Turnbull, Graham A},
  journal   = {Nature Photonics},
  title     = {How to recognize lasing},
  year      = {2009},
  number    = {10},
  pages     = {546--549},
  volume    = {3},
  doi       = {10.1038/nphoton.2009.173},
  publisher = {Nature Publishing Group},
}

@Article{Kreinberg2017,
  author    = {Kreinberg, S{\"o}ren and Chow, Weng W and Wolters, Janik and Schneider, Christian and Gies, Christopher and Jahnke, Frank and H{\"o}fling, Sven and Kamp, Martin and Reitzenstein, Stephan},
  title     = {Emission from quantum-dot high-$\beta$ microcavities: transition from spontaneous emission to lasing and the effects of superradiant emitter coupling},
  journal   = {Light: Science \& Applications},
  year      = {2017},
  volume    = {6},
  number    = {8},
  pages     = {e17030–e17030},
  publisher = {Nature Publishing Group},
  doi={10.1038/lsa.2017.30}
}

@article{KoulasSimos2022,
  doi = {10.1002/lpor.202200086},
  year = {2022},
  month = jul,
  publisher = {Wiley},
  volume = {16},
  number = {9},
  pages = {2200086}, 
  author = {Aris Koulas-Simos and Joel Buchgeister and Monty L. Drechsler and Taiping Zhang and Kaisa Laiho and Georgios Sinatkas and Jialu Xu and Frederik Lohof and Qiang Kan and Ruikang K. Zhang and Frank Jahnke and Christopher Gies and Weng W. Chow and Cun-Zheng Ning and Stephan Reitzenstein},
  title = {Quantum Fluctuations and Lineshape Anomaly in a High-$\beta$ Silver-Coated {InP}-Based Metallic Nanolaser},
  journal = {Laser {\&} Photonics Reviews}
}

@misc{Burger2008,
  author       = {Burger, Sven and Zschiedrich, Lin and Pomplun, Jan and Schmidt, Frank},
  title        = {{JCMsuite}: An Adaptive {FEM} Solver for Precise Simulations in Nano-Optics},
  year         = {2008},
  howpublished = {Integrated Photonics and Nanophotonics Research and Applications},
  note         = {Paper ITuE4},
  publisher    = {Optica Publishing Group},
  address      = {Boston, Massachusetts, USA},
  doi          = {10.1364/IPNRA.2008.ITuE4}
}

@article{Ross2013,
  title={Electrical control of neutral and charged excitons in a monolayer semiconductor},
  author={Ross, Jason S and Wu, Sanfeng and Yu, Hongyi and Ghimire, Nirmal J and Jones, Aaron M and Aivazian, Grant and Yan, Jiaqiang and Mandrus, David G and Xiao, Di and Yao, Wang and Xu, Xiaodong},
  journal={Nature Communications},
  volume={4},
  number={1},
  pages={1474},
  year={2013},
  publisher={Nature Publishing Group UK London},
  doi={10.1038/ncomms2498}
}

@article{Robert2016,
  title={Exciton radiative lifetime in transition metal dichalcogenide monolayers},
  author={Robert, C{\'e}dric and Lagarde, David and Cadiz, Fabian and Wang, Gang and Lassagne, Benjamin and Amand, Thierry and Balocchi, Andrea and Renucci, Pierre and Tongay, Seffattin and Urbaszek, Bernhard and Marie, Xavier},
  journal={Physical Review B},
  volume={93},
  number={20},
  pages={205423},
  year={2016},
  publisher={APS},
  doi={10.1103/PhysRevB.93.205423}
}

@article{Palekar2024,
  title={Enhancement of Interlayer Exciton Emission in a {TMDC} Heterostructure via a Multi-Resonant Chirped Microresonator Upto Room Temperature},
  author={Palekar, Chirag C and Rosa, Barbara and Heermeier, Niels and Shih, Ching-Wen and Limame, Imad and Koulas-Simos, Aris and Rahimi-Iman, Arash and Reitzenstein, Stephan},
  journal={Advanced Materials},
  volume={36},
  number={35},
  pages={2402624},
  year={2024},
  publisher={Wiley Online Library},
  doi={10.1002/adma.202402624}
}

@article{Quan2011,
  title={Deterministic design of wavelength scale, ultra-high {Q} photonic crystal nanobeam cavities},
  author={Quan, Qimin and Loncar, Marko},
  journal={Optics Express},
  volume={19},
  number={19},
  pages={18529--18542},
  year={2011},
  publisher={Optical Society of America},
  doi={10.1364/OE.19.018529}
}

@article{Munkhbat2020,
  title={Transition metal dichalcogenide metamaterials with atomic precision},
  author={Munkhbat, Battulga and Yankovich, Andrew B and Baranov, Denis G and Verre, Ruggero and Olsson, Eva and Shegai, Timur O},
  journal={Nature Communications},
  volume={11},
  number={1},
  pages={4604},
  year={2020},
  publisher={Nature Publishing Group UK London},
  doi= {10.1038/s41467-020-18428-2}
}

@article{Liu2020temperature,
  title={Temperature-dependent optical constants of monolayer {MoS$_2$, MoSe$_2$, WS$_2$, and WSe$_2$}: spectroscopic ellipsometry and first-principles calculations},
  author={Liu, Hsiang-Lin and Yang, Teng and Chen, Jyun-Han and Chen, Hsiao-Wen and Guo, Huaihong and Saito, Riichiro and Li, Ming-Yang and Li, Lain-Jong},
  journal={Scientific Reports},
  volume={10},
  number={1},
  pages={15282},
  year={2020},
  doi={10.1038/s41598-020-71808-y},
  publisher={Nature Publishing Group UK London}
}

@Article{Chow2018,
  author    = {Chow, Weng W and Reitzenstein, Stephan},
  title     = {Quantum-optical influences in optoelectronics—an introduction},
  journal   = {Applied Physics Reviews},
  year      = {2018},
  volume    = {5},
  number    = {4},
  pages     = {041302},
  publisher = {AIP Publishing LLC},
  doi={10.1063/1.5045580}
}

@article{Bjork1991,
  title = {Analysis of semiconductor microcavity lasers using rate equations},
  volume = {27},
  ISSN = {0018-9197},
  DOI = {10.1109/3.100877},
  number = {11},
  journal = {IEEE Journal of Quantum Electronics},
  publisher = {Institute of Electrical and Electronics Engineers (IEEE)},
  author = {Bjork,  G. and Yamamoto,  Y.},
  year = {1991},
  pages = {2386–2396}
}

@article{Kim2016,
  title={Coherent polariton laser},
  author={Kim, Seonghoon and Zhang, Bo and Wang, Zhaorong and Fischer, Julian and Brodbeck, Sebastian and Kamp, Martin and Schneider, Christian and H{\"o}fling, Sven and Deng, Hui},
  journal={Physical Review X},
  volume={6},
  number={1},
  pages={011026},
  year={2016},
  publisher={APS},
  doi={PhysRevX.6.011026}
}

@article{Pan2016,
  title={Dynamic hysteresis in a coherent high-$\beta$ nanolaser},
  author={Pan, Si Hui and Gu, Qing and El Amili, Abdelkrim and Vallini, Felipe and Fainman, Yeshaiahu},
  journal={Optica},
  volume={3},
  number={11},
  pages={1260--1265},
  year={2016},
  publisher={Optical Society of America},
  doi={10.1364/OPTICA.3.001260}
}

@article{Lebreton2015,
  title={Pulse-to-pulse jitter measurement by photon correlation in high-$\beta$ lasers},
  author={Lebreton, Armand and Abram, Izo and Braive, R{\'e}my and Belabas, Nadia and Sagnes, Isabelle and Marsili, Francesco and Verma, Varun B and Nam, Sae Woo and Gerrits, Thomas and Robert-Philip, Isabelle and Stevens, Martin J. and Beveratos, Alexios},
  journal={Applied Physics Letters},
  volume={106},
  number={3},
  pages={031108},
  year={2015},
  publisher={AIP Publishing},
  doi={10.1063/1.4906140}
}

@article{Munkhbat2022optical,
  title={Optical constants of several multilayer transition metal dichalcogenides measured by spectroscopic ellipsometry in the 300--1700 nm range: high index, anisotropy, and hyperbolicity},
  author={Munkhbat, Battulga and Wr{\'o}bel, Piotr and Antosiewicz, Tomasz J and Shegai, Timur O},
  journal={ACS Photonics},
  volume={9},
  number={7},
  pages={2398--2407},
  year={2022},
  publisher={ACS Publications},
  doi={10.1021/acsphotonics.2c00433}
}

@article{Hsu2019,
  title={Thickness-dependent refractive index of {1L}, {2L}, and {3L} {MoS$_2$, MoSe$_2$, WS$_2$, and WSe$_2$}},
  author={Hsu, Chunwei and Frisenda, Riccardo and Schmidt, Robert and Arora, Ashish and De Vasconcellos, Steffen Michaelis and Bratschitsch, Rudolf and van der Zant, Herre SJ and Castellanos-Gomez, Andres},
  journal={Advanced Optical Materials},
  volume={7},
  number={13},
  pages={1900239},
  year={2019},
  publisher={Wiley Online Library},
  doi={10.1002/adom.201900239}
}

@article{Rodriguez2016,
  title={Self-consistent optical constants of {SiO$_2$ and Ta$_2$O$_5$} films},
  author={Rodr{\'\i}guez-de Marcos, Luis V and Larruquert, Juan I and M{\'e}ndez, Jos{\'e} A and Azn{\'a}rez, Jos{\'e} A},
  journal={Optical Materials Express},
  volume={6},
  number={11},
  pages={3622--3637},
  year={2016},
  publisher={Optical Society of America},
  doi={10.1364/OME.6.003622}
}

@article{Fryett2018,
  title={Encapsulated silicon nitride nanobeam cavity for hybrid nanophotonics},
  author={Fryett, Taylor K and Chen, Yueyang and Whitehead, James and Peycke, Zane Matthew and Xu, Xiaodong and Majumdar, Arka},
  journal={ACS Photonics},
  volume={5},
  number={6},
  pages={2176--2181},
  year={2018},
  publisher={ACS Publications},
  doi={10.1021/acsphotonics.8b00036}
}

@article{Qian2024,
  title={Lasing of moir{\'e} trapped {MoSe$_2$/WSe$_2$} interlayer excitons coupled to a nanocavity},
  author={Qian, Chenjiang and Troue, Mirco and Figueiredo, Johannes and Soubelet, Pedro and Villafa{\~n}e, Viviana and Beierlein, Johannes and Klembt, Sebastian and Stier, Andreas V and H{\"o}fling, Sven and Holleitner, Alexander W and Finley, Jonathan J.},
  journal={Science Advances},
  volume={10},
  number={2},
  pages={eadk6359},
  year={2024},
  publisher={American Association for the Advancement of Science},
  doi={10.1126/sciadv.adk6359}
}

@article{Johnson2002perturbation,
  title={Perturbation theory for Maxwell’s equations with shifting material boundaries},
  author={Johnson, Steven G and Ibanescu, Mihai and Skorobogatiy, MA and Weisberg, Ori and Joannopoulos, JD and Fink, Yoel},
  journal={Physical Review E},
  volume={65},
  number={6},
  pages={066611},
  year={2002},
  publisher={APS}
}

@article{Arora2015,
  title={Exciton band structure in layered {MoSe$_2$}: from a monolayer to the bulk limit},
  author={Arora, Ashish and Nogajewski, Karol and Molas, Maciej and Koperski, Maciej and Potemski, Marek},
  journal={Nanoscale},
  volume={7},
  number={48},
  pages={20769--20775},
  year={2015},
  publisher={Royal Society of Chemistry},
  doi={10.1039/C5NR06782K}
}

@article{Asada1984,
  title={Gain and intervalence band absorption in quantum-well lasers},
  author={Asada, Masahiro and Kameyama, Atsushi and Suematsu, Yasuharu},
  journal={IEEE Journal of Quantum Electronics},
  volume={20},
  number={7},
  pages={745–753},
  year={1984},
  publisher={IEEE},
  doi={10.1109/JQE.1984.1072464}
}

@article{Salehzadeh2015,
  doi = {10.1021/acs.nanolett.5b01665},
  year = {2015},
  month = jul,
  publisher = {American Chemical Society ({ACS})},
  volume = {15},
  number = {8},
  pages = {5302–5306},
  author = {Omid Salehzadeh and Mehrdad Djavid and Nhung Hong Tran and Ishiang Shih and Zetian Mi},
  title = {Optically Pumped Two-Dimensional {MoS}$_2$ Lasers Operating at Room-Temperature},
  journal = {Nano Letters}
}

@article{vonHelversen2023,
  doi = {10.1088/2053-1583/acfb20},
  year = {2023},
  month = sep,
  publisher = {{IOP} Publishing},
  volume = {10},
  number = {4},
  pages = {045034},
  author = {Martin von Helversen and Lara Greten and Imad Limame and Ching-Wen Shih and Paul Schlaugat and Carlos Ant{\'{o}}n-Solanas and Christian Schneider and B{\'{a}}rbara Rosa and Andreas Knorr and Stephan Reitzenstein},
  title = {Temperature dependent temporal coherence of metallic-nanoparticle-induced single-photon emitters in a {WSe$_2$} monolayer},
  journal = {2D Materials}
}

\end{document}


\title[Article Title]{Supplementary Information: An integrated all-van der Waals nanobeam laser}


\author*[1]{\fnm{Aris} \sur{Koulas-Simos}}\email{aris.koulas-simos@tu-berlin.de}

\author[2]{\fnm{Pietro} \sur{Metuh}}\email{piemet@dtu.dk}

\author[2]{\fnm{Athanasios} \sur{Paralikis}}\email{athpa@dtu.dk}

\author[1]{\fnm{Kartik} \sur{Gaur}}\email{kartik.gaur@tu-berlin.de}

\author[1]{\fnm{Maximilian} \sur{Klonz}}\email{m.klonz@tu-berlin.de}

\author[1]{\fnm{Imad} \sur{Limame}}\email{imad.limame@tu-berlin.de}

\author[1,3]{\fnm{B\'arbara L. T.} \sur{Rosa}}\email{bltr@unicamp.br}

\author[1]{\fnm{Chirag C.} \sur{Palekar}}\email{c.palekar@tu-berlin.de}

\author*[2]{\fnm{Battulga} \sur{Munkhbat}}\email{bamunk@dtu.dk}

\author*[1]{\fnm{Stephan} \sur{Reitzenstein}}\email{stephan.reitzenstein@tu-berlin.de}

\affil[1]{\orgdiv{Institute for Physics and Astronomy}, \orgname{Technical University of Berlin}, \orgaddress{\street{Hardenbergstr. 36}, \postcode{10623} \city{Berlin}, \country{Germany}}}

\affil[2]{\orgdiv{Department of Electrical and Photonics Engineering}, \orgname{Technical University of Denmark}, \orgaddress{\street{2800 Kgs. Lyngby}, \country{Denmark}}}

\affil[3]{\orgdiv{Institute of Physics “Gleb Wataghin”}, \orgname{State University of Campinas}, \orgaddress{\postcode{13083-859} \city{Campinas}, \country{Brazil}}}

\maketitle

This Supplementary Information provides additional experimental, numerical and theoretical details supporting the results presented in the main text. We first describe the numerical design of the all-van der Waals (vdW) nanobeam cavity, including photonic-band engineering, mode confinement, far-field simulations and benchmarking against previously reported transition-metal dichalcogenide (TMDC) nanocavities. We then present the fabrication and structural characterization of the WS$_2$/MoSe$_2$/WS$_2$ heterostructures, together with an estimate of the cryogenic cavity resonance, optical thickness determination and a statistical analysis of the fabricated nanobeam geometries. Next, we investigate the intrinsic optical properties of the unpatterned heterostructures through temperature-dependent microphotoluminescence (\textmu PL) and reflection spectroscopy, quantify the temperature evolution of the exciton and cavity resonances, extract the effective thermo-optic response of multilayer WS$_2$, and present wavelength-selective real-space measurements confirming cavity-mediated light guiding. Furthermore, we describe the laser rate-equation model used to analyze the excitation-power-dependent characteristics and present complementary streak-camera and photon-correlation measurements that examine the ultra-fast emission dynamics, temporal broadening and fluctuation-driven behavior of the nanobeam lasers. Finally, reproducibility measurements on an independent device further support the observed high-$\beta$ coherent-emission behavior of the all-vdW nanobeam platform.

\section{Numerical simulations: Bandgap engineering, all-vdW nanobeam structure and far field}\label{sec:Numericalsimu} 

To guide the design of the all-vdW nanobeam cavity, three-dimensional finite-element-method (FEM) simulations were performed using JCMsuite \cite{Burger2008}. The simulations employed the geometrical parameters listed in Supplementary Table~\ref{tab1:Simus} together with room-temperature optical constants for all constituent materials \cite{Munkhbat2022optical,Hsu2019,Rodriguez2016}. The cavity was designed to achieve spectral resonance at cryogenic temperatures with the A-exciton of the embedded MoSe$_2$ monolayer (ML) \cite{Ross2013,Robert2016,Palekar2024}. Since reliable low-temperature dielectric functions for the TMDC layers are not available, the target room-temperature resonance wavelength was chosen to be approximately 765~nm, anticipating the moderate ($\approx 6.5$~nm) blueshift of the cavity resonance upon cooling, estimated in Supplementary Section~\ref{sec:cryo-estimation}.

\begin{figure}[!t]
\centering
\includegraphics[width=\textwidth]{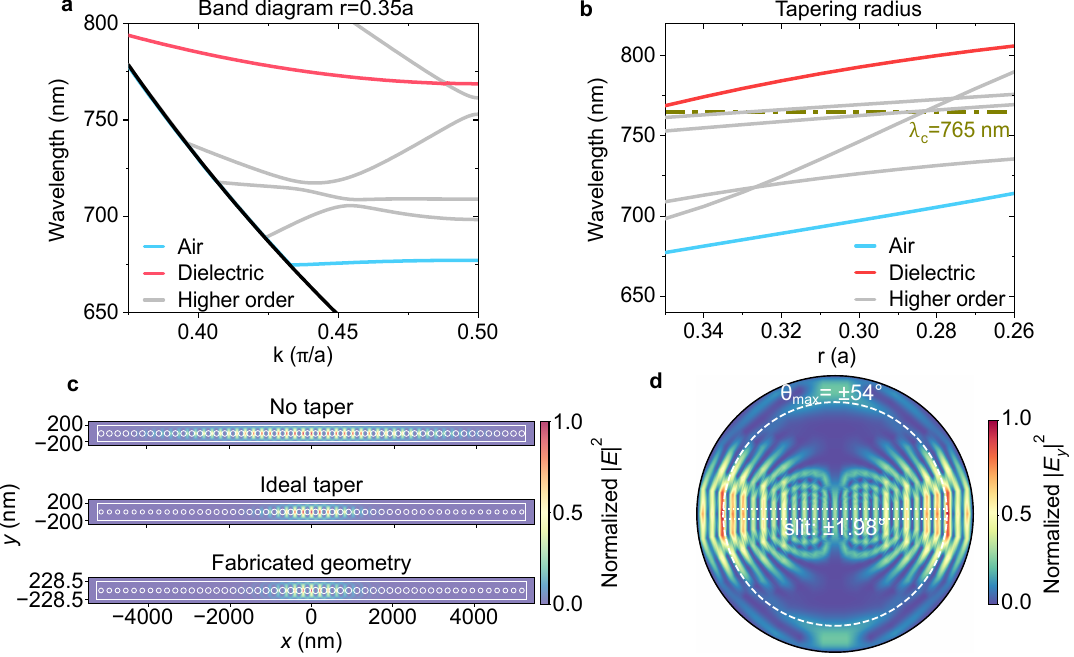}
\caption{\textbf{Numerical simulations of the all-vdW nanobeam cavity.} \textbf{a} Photonic band structure of the unit cell with geometry corresponding to the cavity center $r=r_{\mathrm{in}}$. The red and blue curves denote the dielectric and air band of identical vertical symmetry, respectively, while higher-order bands are shown in gray. \textbf{b} Simulated resonance wavelength of the dielectric band as a function of hole radius, illustrating the bandgap engineering used to position the cavity mode within the photonic bandgap. \textbf{c} Electric-field intensity distributions of the relevant cavity mode for an untapered cavity, the nominal tapered design, and the experimentally realized geometry. The taper produces a near-Gaussian mode profile and significantly improves optical confinement. \textbf{d} Simulated far-field emission profile of the cavity mode. The emission is concentrated at large in-plane momenta, resulting in the strongest collected signal near the maximum collection angle of the objective, in agreement with the experimental observations.}\label{fig:FigureS1}
\end{figure}

The cavity design is based on the dielectric photonic band of a one-dimensional photonic-crystal nanobeam. Supplementary Fig.~\ref{fig:FigureS1}a shows the calculated photonic band structure for the nominal unit-cell geometry (lattice constant $a=200$~nm, width $w=400$~nm, thickness $t_{\mathrm{WS_2}}=274$~nm and hole radius $r_{\mathrm{in}}=70$~nm). The dielectric band (red) and air band (blue) possess identical vertical parity with respect to the nanobeam midplane and therefore can both be confined through a local perturbation of the photonic crystal lattice. Since the electric field of the dielectric band is predominantly concentrated within the WS$_2$ nanobeam, it provides substantially stronger overlap with the embedded MoSe$_2$ ML and was therefore selected for the cavity design. Additional higher-order Bloch bands are shown in gray for completeness. The dielectric band is further characterized by the optical confinement factor
\begin{align}
\Gamma_{\mathrm{WS_2}}
=
\frac{\int_{V_{\mathrm{WS_2}}}\varepsilon(\mathbf r)|\mathbf E(\mathbf r)|^2\,dV}
{\int_{V_{\mathrm{tot}}}\varepsilon(\mathbf r)|\mathbf E(\mathbf r)|^2\,dV},
\end{align}
which quantifies the fraction of the electromagnetic energy confined within the multilayer WS$_2$ nanobeam. For the nominal cavity geometry, approximately 93\% of the optical energy resides inside the dielectric, providing excellent overlap between the confined cavity field and the embedded MoSe$_2$ ML.

To position the cavity resonance within the photonic bandgap, the hole radius was varied systematically while monitoring the wavelength of the dielectric band. The resulting band evolution is shown in Supplementary Fig.~\ref{fig:FigureS1}b. Reducing the hole radius shifts the desired resonance mode deeper into the photonic bandgap. For the present geometry, strong confinement is obtained for $r\approx0.26a$, where the cavity resonance is located close to the center of the bandgap.

The corresponding cavity modes are shown in Supplementary Fig.~\ref{fig:FigureS1}c for three representative geometries: an untapered cavity with $r_{\mathrm{in}}=r_{\mathrm{out}}=70$~nm, the nominal tapered design with $r_{\mathrm{in}}=70$~nm and $r_{\mathrm{out}}=52$~nm, and the experimentally realized geometry using the dimensions summarized in Supplementary Table~\ref{tab1:Simus}. In the tapered structures, the hole radii decrease quadratically from the cavity center towards the mirror region. This profile produces an approximately linear increase in mirror strength from cell to cell, suppressing scattering losses associated with abrupt impedance changes \cite{Quan2011}. Consequently, the cavity mode develops a near-Gaussian envelope along the nanobeam axis, characteristic of adiabatic confinement in photonic-crystal cavities. Compared with the untapered geometry, the taper substantially increases the cavity quality factor (Q-factor) while simultaneously reducing the optical mode volume. The effective mode volume was calculated as
\begin{align}
V_m =
\frac{\int \varepsilon(\mathbf r)|\mathbf E(\mathbf r)|^2\,dV}
{\max\left[\varepsilon(\mathbf r)|\mathbf E(\mathbf r)|^2\right]},
\end{align}
yielding a reduction by approximately a factor of three. The resulting enhancement of light-matter interaction was quantified through the Purcell factor, calculated using a resonance expansion formalism based on the cavity quasinormal modes following Ref.~\cite{Binkowski2025}. This approach accounts for the radiative and lossy nature of the nanobeam cavity and yields the resonance wavelengths, Q-factors, mode volumes, confinement factors and Purcell enhancements summarized in Supplementary Table~\ref{tab1:Simus}.
\begin{table}[!b]
    \centering
    \caption{Relevant simulated optical quantities for a nanobeam without taper, with taper and fabricated case.}
         \begin{tabular}{c c c c c}
            \hline
            \bf{Structure} & \bf{Without taper} & \bf{With taper (nominal)} &  \bf{Fabricated} \\
            \hline   
            Lattice constant $a$ (nm)                        &  $200$       & $200$      & $203$        \\
            Width  $w$ (nm)                                  &  $400$       & $400$      & $457$        \\
            Thickness  $t$ (nm)                              &  $274$       & $274$      & $274$        \\
            Inner radius  $r_{\mathrm{in}}$ (nm)             &  $70$        & $70$       & $80$         \\
            Outer radius  $r_{\mathrm{out}}$ (nm)            &  $70$        & $52$       & $59$         \\\\
            Cavity wavelength $\lambda_c$ (nm)               &  $769$       & $771$      & $763$        \\
            Q-factor                                         &  $8194$      & $14493$    & $13670$      \\
            Mode volume $V_m$ $(\lambda_c/\mathrm{n})^3$     &  $6.84$      & $2.20$     & $2.34$       \\
            Confinement factor $\Gamma_{\mathrm{ML}}$        &  $0.84\%$    & $0.84\%$   & $0.80\%$     \\
            Purcell factor $F_P$                             &  $26$        & $143$      & $125$        \\
            \hline
        \end{tabular}
    \label{tab1:Simus}
\end{table}

The calculated modal properties reveal a clear advantage of the tapered cavity over the untapered geometry. Importantly, the experimentally realized cavity continues to support a strongly confined dielectric-band mode despite the systematic fabrication-induced variations in nanobeam width and hole diameter discussed in Section~\ref{sec:Fabricationquality}. Increasing the nanobeam width alone would redshift the cavity resonance through a larger effective refractive index, whereas the enlarged air holes reduce the dielectric filling fraction and produce a stronger blueshift. As shown in Supplementary Table~\ref{tab1:Simus}, the air-hole-induced blueshift dominates, resulting in a net shift of the cavity resonance towards shorter wavelengths. The simulated resonance wavelength of the fabricated geometry, $\lambda_{\mathrm c}\approx763$~nm at room temperature, is furthermore in excellent agreement with the cavity wavelength obtained by extrapolating the experimentally measured temperature-dependent cavity shift to room temperature, providing independent validation of the numerical model (Supplementary Section~\ref{sec:thermo-optic}). Despite the fabrication-induced dimensional variations, the calculated optical properties of the fabricated cavity remain remarkably close to those of the optimized tapered design and differ substantially less than those of the untapered cavity. This demonstrates the robustness of the taper concept against moderate fabrication deviations.

Finally, the far-field emission profile of the cavity mode was calculated to facilitate comparison with the Fourier-space measurements presented in Fig.~2d of the main text. The simulated far-field intensity distribution is shown in Supplementary Fig.~\ref{fig:FigureS1}d. Consistent with its origin from a guided dielectric-band mode, the emission is concentrated at large in-plane momenta rather than around normal incidence. To reproduce the experimental configuration, the finite acceptance window of the monochromator entrance slit was included in the simulations. For the experimental slit width of 200~\textmu m along the $y$-direction, the calculated emission is strongest near the maximum collection angle of the objective ($\theta\approx54^\circ$), in excellent agreement with the measured Fourier-space distribution. This agreement confirms both the waveguide-derived nature of the cavity mode and the interpretation of the directional emission presented in the main text.

To benchmark the optical performance of the presented all-vdW nanobeam cavity, Supplementary Table~\ref{tab2:Comparison1} compares representative hybrid TMDC photonic-crystal cavities and nanobeam resonators employing transferred TMDC MLs or heterobilayers as the gain medium. Although dielectric platforms based on GaP, Si and SiN generally achieve substantially higher intrinsic Q-factors owing to their lower optical losses, the present all-vdW cavity exhibits a competitive mode volume together with a considerably larger ML confinement factor. Unlike hybrid architectures, where the active ML interacts only with the evanescent tail of the cavity field, the embedded MoSe$_2$ ML is located directly within the optical resonator, yielding $\Gamma_{\mathrm{ML}}=0.8\%$, which significantly exceeds the values of 0.0145$\%$ reported for previous hybrid TMDC cavity platforms \cite{Li2017} and underscores the superiority of the integrated all-vdW concept. The experimentally measured Q-factor corresponds to the loaded cavity Q-factor extracted from PL measurements and therefore includes additional losses arising from material absorption and scattering. Consequently, it should not be directly compared with intrinsic cavity Q-factors obtained from numerical simulations. Together, the combination of a small mode volume, large simulated Purcell enhancement and an embedded gain medium establishes the presented cavity as a competitive platform for enhanced light-matter interaction and coherent emission in fully integrated vdW photonic systems.
\begin{table}[!h]
    \centering
    \caption{Comparison of the simulated and experimentally measured optical properties of representative hybrid TMDC nanobeam and photonic-crystal cavities reported in the literature with the all-vdW nanobeam cavity presented in this work. Here, $Q_\mathrm{exp}^{\mathrm{loaded}}$ denotes the experimentally measured loaded cavity Q-factor after integration of the TMDC gain medium.}
         \begin{tabular}{c c c c c }
            \hline
            \bf{Platform} & $\mathbf{Q_\mathrm{sim}}$ & $\mathbf{Q^{\mathrm{loaded}}_\mathrm{exp}}$ & $\mathbf{V_m}$ $(\lambda_c/\mathrm{n})^3$ & $\mathbf{\Gamma_{\mathrm{ML}}} (\%)$\\
            \hline   
            WSe$_2$-ML on GaP-PCC (L3) \cite{Wu2015}      & n. r.     & $2500$   & $\approx 1$ & n. r.  \\
            MoTe$_2$-ML on Si-NB \cite{Li2017}            & $6.5\times10^5$      & $2830$   & $0.48$  & $0.0145$     \\
            WSe$_2$-ML on SiN-NB \cite{Fryett2018}        & $\sim10^5$      & $320-830$   & $2.5$  & n. r.     \\
            WSe$_2$/MoSe$_2$-hBL on SiN-NB \cite{Qian2024}             & n. r.      & $12500$   & $ 1.2$  & n. r.    \\
            \textbf{This work}           & $\mathbf{13670}$      & $\mathbf{210}$   & $\mathbf{2.34}$  & $\mathbf{0.8}$     \\
            \hline
        \end{tabular}
    \label{tab2:Comparison1}
    \footnotesize{\textbf{n.r.}: not reported. The optical confinement factor $\Gamma_{\mathrm{ML}}$ is generally not reported for hybrid TMDC cavity architectures.}
\end{table}
\section{Estimation of the cryogenic cavity resonance wavelength} \label{sec:cryo-estimation}

To estimate the expected temperature-induced resonance shift of the nanobeam cavity, we assume that the resonance wavelength is primarily governed by the refractive index of the WS$_2$ nanobeam. Since no temperature-dependent refractive-index data are available for multilayer WS$_2$ near the cavity wavelength, we employ the room-temperature refractive index reported in Ref. \cite{Munkhbat2022optical}, $n_{300}=4.2057$ at $\lambda=763$~nm, together with the effective room-temperature thermo-optic coefficient reported for ML WS$_2$, $\left(dn/dT\right)_{\mathrm{lit}}\approx1.3\times10^{-4}~\mathrm{K^{-1}}$ \cite{Liu2020temperature}, as a first-order estimate.

The cavity resonance shift is related to a perturbation of the dielectric function through first-order electromagnetic perturbation theory \cite{Johnson2002perturbation},
\begin{equation}
\frac{\Delta\omega}{\omega}
=
-\frac{1}{2}
\frac{\displaystyle\int \Delta\varepsilon(\mathbf{r})|\mathbf{E}(\mathbf{r})|^2\,dV}
{\displaystyle\int \varepsilon(\mathbf{r})|\mathbf{E}(\mathbf{r})|^2\,dV},
\end{equation}
where $\varepsilon=n^2$ and $\mathbf{E}$ is the electric field of the unperturbed cavity mode. Assuming that only the refractive index of the WS$_2$ region varies with temperature and accounting for first-order perturbations $\Delta\varepsilon = 2n\,\Delta n$, the resonance shift becomes
\begin{equation}
\frac{\Delta\lambda}{\lambda}
=
\Gamma_{\mathrm{WS_2}}
\frac{\Delta n}{n},
\end{equation}
where $\Gamma_{\mathrm{WS_2}}$ is the electric-field energy confinement factor of the cavity mode inside the WS$_2$ nanobeam. Finite-element simulations yield $\Gamma_{\mathrm{WS_2}}=0.93$, indicating that 93\% of the stored electric-field energy resides within the WS$_2$ region. Assuming a temperature-independent thermo-optic coefficient, the refractive-index variation is approximated as
\begin{equation}
\Delta n
=
\left(\frac{dn}{dT}\right)_{\mathrm{lit}}
(T-T_{300}),
\end{equation}
which leads to the expected cavity resonance
\begin{equation}
\lambda(T)
=
\lambda_{300}
\left[
1+
\frac{\Gamma}{n_{300}}
\left(\frac{dn}{dT}\right)_{\mathrm{lit}}
(T-T_{300})
\right].
\end{equation}
Using $\lambda_{300}=763$~nm, $n_{300}=4.2057$, $\Gamma_{\mathrm{WS_2}}=0.93$, and $\left(dn/dT\right)_{\mathrm{lit}}=1.3\times10^{-4}~\mathrm{K^{-1}}$, the estimated cavity wavelength at 4~K is
\begin{equation}
\lambda(4~\mathrm{K})
\approx
756.5~\mathrm{nm},
\end{equation}
corresponding to an expected blueshift of approximately $6.5$~nm upon cooling from room temperature to cryogenic temperature. This estimate is in excellent agreement with the experimentally observed low-temperature cavity resonance near $756.5$~nm, as discussed in Supplementary Section~\ref{sec:thermo-optic}.

\section{Heterostructure fabrication and optical characterization} \label{sec:opticalmicroscopHS}

The WS$_2$/MoSe$_2$/WS$_2$ heterostructures were assembled by deterministic dry-transfer stacking of mechanically exfoliated flakes, followed by electron-beam lithography (EBL) and inductively coupled plasma reactive-ion etching (ICP-RIE) to define the photonic-crystal nanobeam cavities \cite{Metuh2025}. Prior to nanofabrication, the thicknesses of the top and bottom WS$_2$ multilayer flakes were determined from broadband optical reflectivity measurements. The measured spectra were fitted using a transfer-matrix model employing the room-temperature dielectric functions of WS$_2$. As shown in Supplementary Fig.~\ref{fig:FigureS2}a, b, the calculated spectra reproduce the experimental reflectivity with excellent agreement, yielding thicknesses of 132~nm and 137~nm for the top and bottom WS$_2$ layers, respectively. Supplementary Fig.~\ref{fig:FigureS2}c shows an optical microscope image of a representative fabricated heterostructure after nanobeam patterning. The dashed outline indicates the position of the embedded MoSe$_2$ ML, while the highlighted nanobeam cavities (NB1 and NB2) correspond to the devices investigated throughout the main text and Supplementary Information.

\begin{figure}[!t]
\centering
\includegraphics[width=\textwidth]{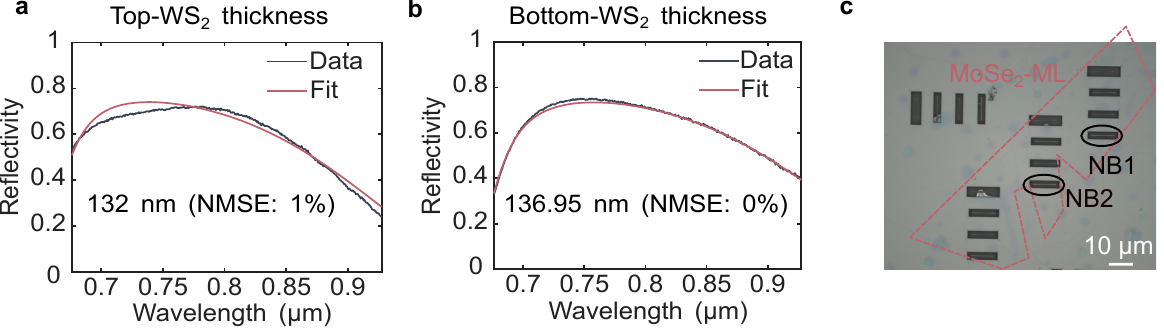}
\caption{\textbf{Heterostructure characterization.} \textbf{a, b} Broadband optical reflectivity spectra (gray) of the top and bottom WS$_2$ flakes together with transfer-matrix fits (red), yielding thicknesses of 132~nm and 137~nm, respectively. \textbf{c} Optical microscope image of the fabricated WS$_2$/MoSe$_2$/WS$_2$ heterostructure after nanobeam fabrication. The dashed outline marks the embedded MoSe$_2$ ML, while the highlighted nanobeams correspond to the devices investigated in the main text.}\label{fig:FigureS2}
\end{figure}

\section{Fabrication quality: SEM analysis of nanobeam geometries}\label{sec:Fabricationquality}
To assess the fabrication accuracy and reproducibility of the nanobeam resonators, a statistical geometrical analysis was performed using scanning electron microscopy (SEM). The analyzed structures were fabricated from a single 270~nm-thick WS$_2$ flake using the same transfer \cite{Metuh2025}, EBL and ICP-RIE procedures as employed for the optically active devices investigated in the main text. Representative SEM images are shown in Supplementary Fig.~\ref{fig:FigureS3}a-c. The fabricated nanobeams exhibit well-defined photonic-crystal features with a high degree of structural uniformity along the entire cavity length. A slight lateral anisotropy of the etched holes results in rounded hexagonal apertures rather than perfectly circular ones \cite{Munkhbat2020}. This morphology is attributed to the anisotropic dry-etching characteristics of multilayer WS$_2$ and is consistently observed across multiple fabricated devices.

\begin{figure}[!t]
\centering
\includegraphics[width=\textwidth]{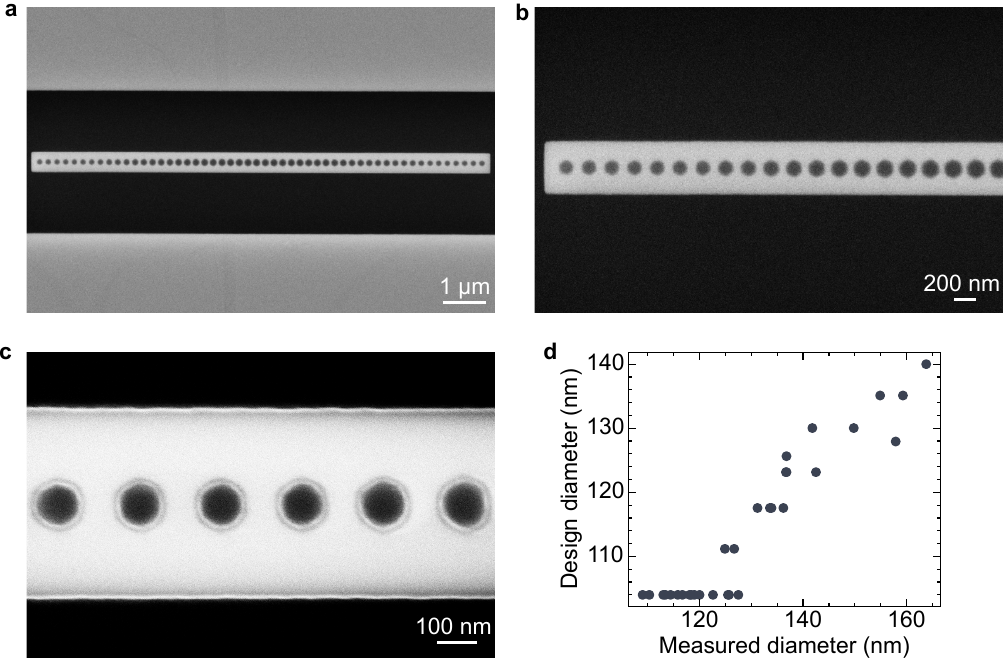}
\caption{\textbf{SEM on WS$_2$ nanobeam cavities}. \textbf{a} Image of one of the fabricated nanobeam cavities, patterned from a 270~nm-thick WS$_2$ flake following the same process as the optically active devices in the main text. \textbf{b} Magnified image of the same cavity, showing the small lateral anisotropy in the etched holes, which have a rounded hexagonal shape. \textbf{c} Example of high-resolution image used for the geometrical analysis of the nanobeam cavities. \textbf{d} Deviation of the measured hole diameter size compared to their design size.}\label{fig:FigureS3}
\end{figure}
The lattice constant was determined from the center-to-center spacing of $n=49$ consecutive holes within the cavity region, yielding an average value of $a=(203.11\pm2.17)$~nm. The extracted standard deviation is only slightly larger than the SEM pixel resolution of approximately 1.6~nm, demonstrating excellent control over the hole placement during fabrication.

The nanobeam width was extracted from four independent positions along the cavity and yielded an average value of $w=(457.7\pm8.3)$~nm, corresponding to an increase of approximately $14.4\%$ relative to the nominal design. 

To further quantify the fabrication accuracy, the diameters of all patterned holes were extracted from high-resolution SEM images, with a representative one shown in Supplementary Fig.~\ref{fig:FigureS3}c. The measured diameters are compared with their nominal design values in Supplementary Fig.~\ref{fig:FigureS3}d. Across $n=37$ analyzed holes, the average diameter exceeds the design value by $(15.6\,\pm\,5.4)$~nm, corresponding to a relative increase of $(13.8\,\pm\,4.4)\%$. This closely matches the increase observed for the nanobeam width, indicating that the fabrication process introduces an approximately uniform lateral scaling rather than significant local distortions. The small spread of the extracted dimensions further demonstrates the reproducibility of the nanofabrication process and justifies the use of the experimentally determined geometrical parameters in the numerical simulations presented in Supplementary Section~\ref{sec:Numericalsimu}.

\section{Temperature-dependent \texorpdfstring{\textmu}{µ}PL measurements on unpatterned
\texorpdfstring{WS$_2$/MoSe$_2$-ML/WS$_2$}{WS2/MoSe2-ML/WS2} heterostructure region and comparison with nanobeam} \label{sec:unpatterned}
To establish the intrinsic optical properties of the embedded MoSe$_2$ ML and provide a reference for the cavity measurements discussed in the main text, temperature-dependent \textmu PL and \textmu reflection spectroscopy were performed on an unpatterned region of the same WS$_2$/MoSe$_2$/WS$_2$ heterostructure (inset in Supplementary Fig.~\ref{fig:FigureS4}b). The results are summarized in Supplementary Fig.~\ref{fig:FigureS4}.

The temperature-dependent \textmu PL spectra are shown in Supplementary Figure~\ref{fig:FigureS4}a. At low temperature, the emission is dominated by a single excitonic transition centered near 761~nm, attributed to the neutral A exciton of the MoSe$_2$ ML. In contrast to many previously reported exfoliated MoSe$_2$ MLs \cite{Ross2013,Robert2016}, no distinct trion emission is observed. The absence of a measurable charged-exciton contribution indicates low residual doping and reduced charge disorder within the encapsulated heterostructure, consistent with the high optical quality required for efficient exciton-cavity coupling.

Upon increasing temperature, the excitonic emission continuously redshifts, broadens and decreases in intensity, consistent with bandgap renormalization and enhanced exciton-phonon scattering \cite{Ross2013,Arora2015}. The extracted integrated intensity, emission wavelength and linewidth are summarized in Supplementary Fig.~\ref{fig:FigureS4}b--d. Between 10 and 180~K, the exciton redshifts by approximately 20~nm, whereas the linewidth increases from about 3 to more than 12~nm. Over the same temperature range, the integrated emission intensity decreases by approximately one order of magnitude owing to increasingly efficient non-radiative recombination. The substantially larger thermal shift of the exciton compared with the cavity mode enables controlled tuning of the exciton-cavity detuning, as exploited in the main text.
\begin{figure}[!t]
\centering
\includegraphics[width=\textwidth]{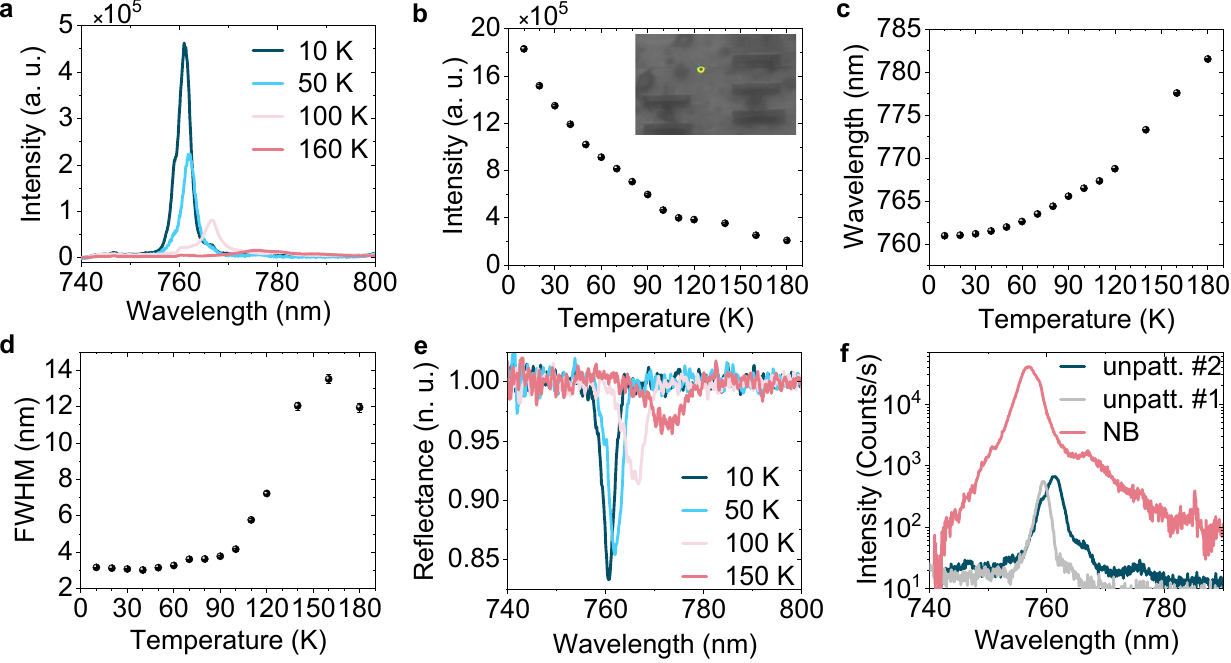}
\caption{\textbf{Temperature-dependent optical properties of the unpatterned WS$_2$/MoSe$_2$/WS$_2$ heterostructure.} \textbf{a} Temperature-dependent \textmu PL spectra measured between 10~K and 180~K. \textbf{b-d} Extracted integrated intensity, exciton wavelength and linewidth as a function of temperature. Optical microscope of a representative excitation region shown in \textbf{b} as inset. \textbf{e} Temperature-dependent \textmu reflection spectra confirming the excitonic resonance and its spectral evolution. \textbf{f} Comparison of low-temperature PL spectra recorded from the nanobeam cavity and two unpatterned regions of the heterostructure, revealing an approximately 70-fold enhancement of the cavity emission intensity.}\label{fig:FigureS4}
\end{figure}
To independently confirm the excitonic origin of the observed transition, temperature-dependent \textmu reflection measurements were performed on the same heterostructure. A nearby multilayer region without the embedded ML was used as reference measurement. The baseline-corrected and smoothed spectra are shown in Supplementary Fig.~\ref{fig:FigureS4}e. A pronounced excitonic absorption resonance is observed over the entire investigated temperature range and follows the same thermal redshift as the PL spectra, confirming that the measured transition originates from the intrinsic MoSe$_2$ exciton.

Finally, Supplementary Fig.~\ref{fig:FigureS4}f compares low-temperature PL spectra recorded from the nanobeam cavity and two representative unpatterned regions of the heterostructure. While the unpatterned regions exhibit the excitonic emission characteristic of the embedded MoSe$_2$ ML, the nanobeam cavity shows an approximately 70-fold higher emission intensity at the spectral resonance. This enhancement demonstrates efficient funneling of spontaneous emission into the confined cavity mode and confirms that the pronounced cavity emission observed in the main text originates from optical confinement rather than local variations in material quality.

\section{Temperature dependence of the exciton and cavity resonances and extraction of the thermo-optic response of multilayer \texorpdfstring{WS$_2$}{WS2}}\label{sec:thermo-optic}

\begin{figure}[!t]
\centering
\includegraphics[width=\textwidth]{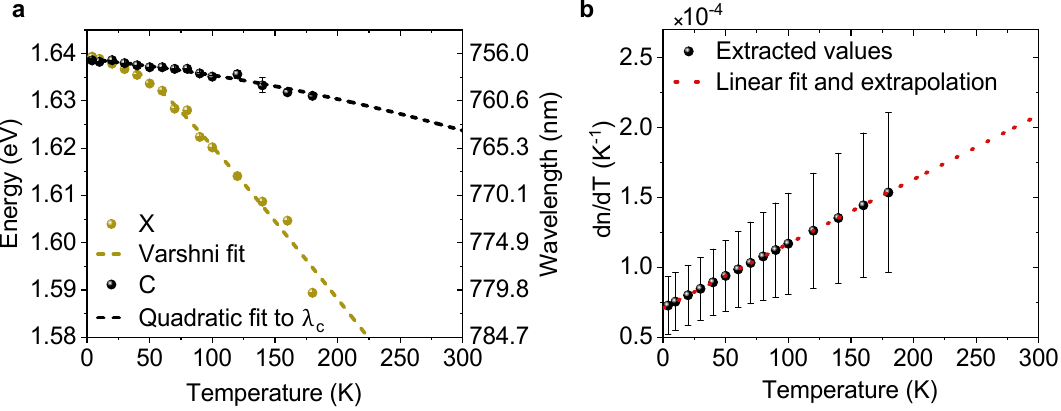}
\caption{\textbf{Temperature dependence of the exciton (X) and cavity (C) resonances and extraction of the effective thermo-optic coefficient.} \textbf{a} Temperature-dependent neutral MoSe$_2$ A-exciton and nanobeam cavity resonances extracted from the PL spectra. The exciton resonance is shown by the gold symbols and the cavity resonance by the black symbols. The gold dashed curve is a fit of the exciton energy to the modified Varshni (Bose--Einstein) model following Ref.~\cite{Ross2013}. The black curve is an empirical second-order polynomial fit to the temperature-dependent cavity wavelength. \textbf{b} Effective thermo-optic coefficient of the multilayer WS$_2$ nanobeam obtained from the cavity-resonance shift using Eq.~(\ref{eq:thermooptic_extraction}). Black symbols show the values extracted at the measured temperatures, with error bars representing propagated uncertainties from the cavity-wavelength extraction and the polynomial-fit parameters. The red line is an unweighted linear fit over the measured range of 4--180~K and serves as a guide to the observed trend, while the values above 180 K are extrapolated.}
\label{fig:FigureS5}
\end{figure}

To quantify the temperature-dependent exciton-cavity detuning shown in Fig.~2a of the main text, the emission wavelengths of the neutral MoSe$_2$ A-exciton and the nanobeam cavity mode were extracted from the temperature-dependent PL spectra (Supplementary Fig.~\ref{fig:FigureS5}a). Both resonances redshift with increasing temperature, although the exciton exhibits a substantially stronger temperature dependence than the cavity mode.
The exciton transition energy was fitted using the modified Varshni relation employed for MoSe$_2$ ML in Ref.~\cite{Ross2013},
\begin{equation}\label{eq:modified_varshni}
E_{\mathrm{X}}(T) = E_{\mathrm{X}}(0) -S\langle\hbar\omega\rangle\left[\coth\left(
\frac{\langle\hbar\omega\rangle}{2k_{\mathrm{B}}T}\right) -1\right],
\end{equation}
where $E_{\mathrm{X}}(0)$ is the exciton transition energy at zero temperature, $S$ is a dimensionless exciton-phonon coupling parameter, $\langle\hbar\omega\rangle$ is an effective average phonon energy, and $k_{\mathrm{B}}$ is the Boltzmann constant. The fit yields
\begin{align}
E_{\mathrm{X}}(0)=(1.63846\pm0.00044)~\mathrm{eV},\,S=1.98\pm0.12,\, \langle\hbar\omega\rangle= (10.0\pm1.1)~\mathrm{meV}.
\end{align}
The extracted trend reproduces the pronounced thermal redshift of the exciton and is consistent with temperature-induced band-gap renormalization mediated by exciton-phonon interactions. Over the experimentally investigated range, the exciton shifts from approximately $756.3$~nm at 4~K to $780.1$~nm at 180~K.

In contrast, the cavity mode exhibits a substantially weaker thermal redshift. The measured cavity resonance displays a variable slope and was therefore described by an empirical second-order polynomial
\begin{equation} \label{eq:cavity_quadratic}
\lambda_{\mathrm c}(T)=a+bT+cT^2,
\end{equation}
with the fitted parameters:
\begin{align}
a &= (756.53282\pm0.10599)~\mathrm{nm},\\
b &= (0.01185\pm0.00344)~\mathrm{nm\,K^{-1}},\\
c &= (3.87909\pm2.49668)\times10^{-5}~\mathrm{nm\,K^{-2}}.
\end{align}
The measured cavity shifts can be used to extract an effective thermo-optic coefficient of the multilayer WS$_2$ nanobeam. Differentiating the first-order perturbation relation introduced in the Supplementary Section~\ref{sec:cryo-estimation} gives
\begin{equation} \label{eq:thermooptic_extraction}
\left(\frac{dn}{dT}\right)_{\mathrm{eff}} = \frac{n} {\Gamma_{\mathrm{WS_2}}\lambda_{\mathrm{c}}} \frac{d\lambda_{\mathrm{c}}}{dT}= \frac{n}{\Gamma_{\mathrm{WS_2}}}\frac{b+2cT}{\lambda_{\mathrm c}(T)},
\end{equation}
where $n=4.2057$ is the refractive index used for multilayer WS$_2$ and $\Gamma_{\mathrm{WS_2}}=0.93$ is the optical confinement factor obtained from the electromagnetic simulations. The uncertainties of the extracted thermo-optic coefficients were obtained by propagating the uncertainties of the measured cavity wavelengths and the fitted polynomial coefficients.

The extracted effective thermo-optic coefficients are shown in Fig.~\ref{fig:FigureS5}b. Over the experimentally investigated temperature range, the effective thermo-optic coefficient increases from approximately $7.3\times10^{-5}$~K$^{-1}$ at 4~K to approximately $1.5\times10^{-4}$~K$^{-1}$ at 180~K. The extracted values are approximately described by a linear temperature dependence over this restricted range, and the linear fit shown in Fig.~\ref{fig:FigureS5}b is therefore included as a guide to the observed experimental trend rather than as a microscopic model.

For comparison with room-temperature values, the empirical quadratic interpolation was evaluated at 300~K. This gives an extrapolated cavity wavelength of $\lambda_{\mathrm c}(300~\mathrm{K}) =763.58~\mathrm{nm}$ in excellent agreement with the simulated resonance wavelength at room temperature and a thermo-optic coefficient $\left.\left(\frac{dn}{dT}\right)_{\mathrm{eff}} \right|_{300\,\mathrm K} \approx 2.08\times10^{-4}~\mathrm{K^{-1}}$. Because the measurements extend only to 180~K, both the cavity wavelength and the effective thermo-optic coefficient quoted at 300~K should be regarded as extrapolations of the empirical polynomial and not as independently measured room-temperature values. The extrapolated coefficient is of the same order of magnitude as previously reported room-temperature thermo-optic coefficients for WS$_2$ \cite{Liu2020temperature}.

The markedly different thermal shifts of the exciton and cavity resonance provide continuous control of their spectral detuning. Whereas the cavity resonance redshifts by approximately 3.5~nm between 4 and 180~K, the exciton resonance shifts by approximately 23.8~nm over the same range. Consequently, their near-resonant spectral alignment at cryogenic temperature evolves into an increasingly large detuning with increasing temperature, consistent with the temperature-dependent spectra presented in Fig.~2a of the main text.

\section{Wavelength-selective real-space imaging of the cavity mode}
\begin{figure}[!t]
\centering
\includegraphics[width=\textwidth]{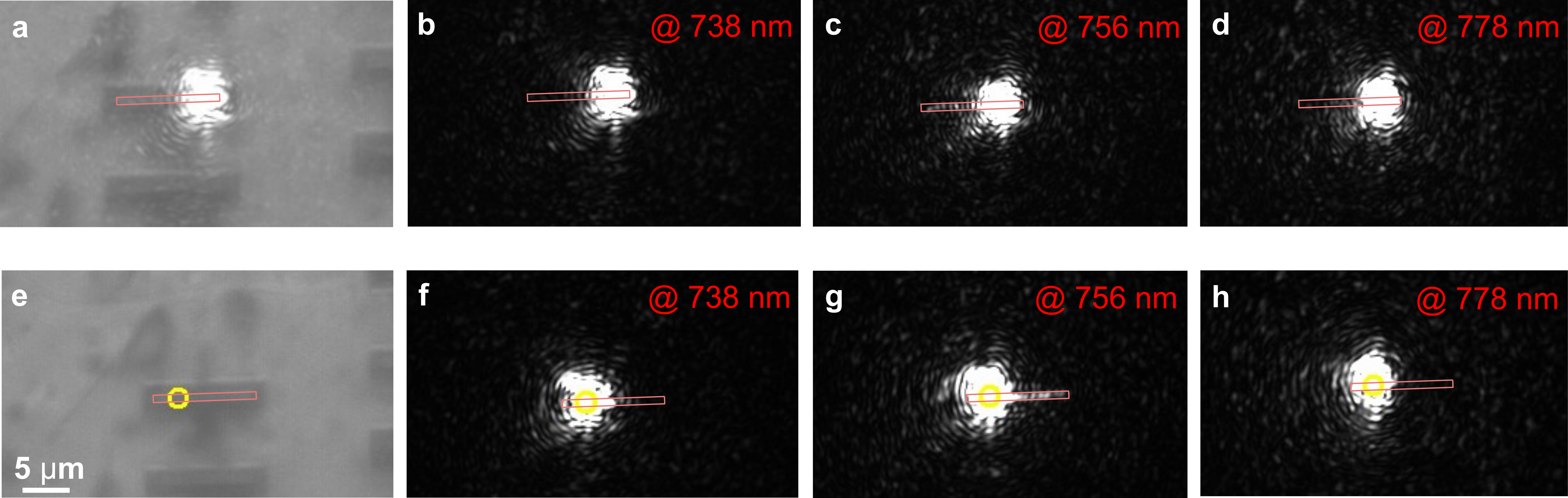}
\caption{\textbf{Wavelength-selective real-space PL imaging.} \textbf{a, e} Bright-field microscope images indicating the excitation position at the left (\textbf{a}) and right (\textbf{e}) end of the nanobeam cavity.  \textbf{b--d} Real-space PL images recorded for excitation at the left end while detecting at wavelengths of 738 nm, 756 nm and 778~nm, respectively. \textbf{f--h} Corresponding measurements for excitation at the right end of the nanobeam. Pronounced emission extending along the nanobeam is observed only at the cavity resonance wavelength (756~nm), whereas off-resonant detection at 738 nm and 778~nm yields only localized emission around the excitation spot. These measurements confirm that the elongated emission observed in the main text originates from coupling to the nanobeam cavity mode.}\label{fig:FigureS6}
\end{figure}

To verify the cavity-mediated origin of the real-space emission shown in the main text, wavelength-selective PL images were recorded while exciting either the left or right end of the nanobeam cavity. Supplementary Fig.~\ref{fig:FigureS6} presents images acquired for excitation wavelengths of 738 nm, 756 nm and 778~nm.

For both excitation positions, an elongated emission feature extending along the nanobeam is observed only at the cavity resonance wavelength of 756~nm. In contrast, excitation at 738 nm and 778~nm yields only localized emission around the excitation spot due to the reflection of the excitation laser, without any measurable propagation along the nanobeam.

The exclusive appearance of the guided emission at the cavity resonance demonstrates that the observed signal originates from resonant coupling to the fundamental nanobeam cavity mode. The absence of a comparable feature away from resonance rules out broadband excitonic emission or stray scattering as the origin of the elongated emission and further supports the interpretation of the real-space measurements presented in the main text.

\section{Numerical model for laser rate equation fitting}\label{sec:rate-equation}

The excitation-power-dependent emission characteristics of the all-vdW nanobeam cavities were analyzed using a standard high-$\beta$ laser rate-equation model based on a two-level gain system \cite{Bjork1991}. Such models are widely employed to describe the gradual transition from spontaneous to stimulated emission in nanolasers and provide a framework for estimating the spontaneous-emission coupling factor $\beta$ and the threshold pump power from experimentally measured input-output (I/O) characteristics. The analytical solution of the coupled carrier-photon rate equations relates the optical pump power $P_{\mathrm{pump}}$ to the emitted output power $P_{\mathrm{out}}$ according to
\begin{equation}\label{eq:rate-eq}
P_{\text{pump}}\left(A,B,\beta,\gamma\right)=A\frac{\gamma}{\beta} \left[\frac{BP_{\mathrm{out}}}{1+P_{\mathrm{out}}}\left(1+\xi\right)\left(1+\beta BP_{\mathrm{out}}\right)-\beta\xi B P_{\mathrm{out}}\right],
\end{equation}
where $A$ and $B$ are scaling parameters linking the experimentally measured pump and output powers to the carrier and photon populations within the cavity. The dimensionless parameter $\xi=\xi_P\beta = \frac{n_0\beta}{\gamma\tau_{\mathrm{sp}}}$ contains the transparency exciton population $n_0$, the spontaneous-emission lifetime $\tau_{\mathrm{sp}}$, the cavity decay rate $\gamma$ and the spontaneous-emission coupling factor $\beta$. The cavity decay rate is calculated from the measured cavity resonance wavelength $\lambda$ and quality factor $Q$ through $\gamma=\frac{2\pi c}{\lambda Q}$, where $c$ denotes the speed of light. Since $\xi$ depends explicitly on $\beta$, it is recalculated self-consistently during the fitting procedure based on the fixed $\xi_P$ and is not treated as an independent fit parameter.

Within this framework, the threshold pump power is defined as the pump level at which the average intracavity photon number reaches unity and is given by
\begin{equation}\label{eq:rate-thresh}
P_{\mathrm{th}}= A\left[\xi\left(1-\beta\right)+1 +\beta\right]\frac{\gamma}{2\beta}.
\end{equation}

To estimate the transparency exciton population $n_0$, we first evaluate the transparency carrier density $N_0$, defined as the carrier density for which the electron and hole occupation probabilities become equal at the relevant emission energy. Following the quantum-well treatment introduced by Asada \textit{et al.} \cite{Asada1984} and subsequently applied to TMDC ML gain media \cite{Salehzadeh2015,KoulasSimos2024}, the carrier occupations are described by
\begin{align}
f_e(E) &= \left[1+\mathrm{exp}\left(\frac{m_rE}{m_e k_B T}-\mathrm{ln}\left(\mathrm{exp}\left(\frac{N\pi \hbar^2 w}{m_ek_BT}-1\right)\right)\right)\right]^{-1} ,\\
f_h(E) &= \left[1+\mathrm{exp}\left(-\frac{m_rE}{m_h k_B T}+\mathrm{ln}\left(\mathrm{exp}\left(\frac{N\pi \hbar^2 w}{m_hk_BT}-1\right)\right)\right)\right]^{-1} .
\end{align}
Here, $m_e$ and $m_h$ denote the effective electron and hole masses, respectively, $m_r$ is the reduced mass, $w$ is the ML thickness and $N$ the carrier density. Transparency at the relevant emission energy is reached at approximately $N_0=9\times10^{24}~\mathrm{m}^{-3}$, corresponding to a two-dimensional carrier density of $N^{(\mathrm{2D})}_0=6\times10^{11}~\mathrm{cm}^{-2}$. Using the ML thickness and the experimentally determined excitation area, this corresponds to a transparency exciton population of approximately $n_0\approx7.8\times10^2$.

The spontaneous-emission lifetime of excitonic states in TMDC MLs can vary substantially depending on the dielectric environment and localization potential \cite{vonHelversen2023}. For the present analysis, a value of $\tau_{\mathrm{sp}}=1.8$~ps was used, consistent with previously reported low-temperature measurements on MoSe$_2$ MLs \cite{Robert2016}. The cavity resonance wavelength $\lambda_c=756.5$~nm and quality factor $Q=210$ were obtained directly from the experimentally measured spectra and the corresponding Voigt-profile fits. The measured I/O characteristics were fitted using the rate-equation model with $\beta$, $A$ and $B$ as free parameters, while $n_0$, $\tau_{\mathrm{sp}}$ and $\gamma$ were fixed to their independently estimated values. The resulting fits provide estimates of the spontaneous-emission coupling factor and the corresponding threshold pump power of the nanobeam cavities. We note that, owing to the smooth transition between spontaneous and stimulated emission in high-$\beta$ nanolasers, rate-equation analysis alone cannot unambiguously distinguish lasing from amplified spontaneous emission \cite{Kreinberg2017}. The extracted fitting parameters should therefore be interpreted together with the excitation-power-dependent linewidth evolution and, most importantly, the photon-correlation measurements presented in the main text, which provide direct quantum-optical verification of coherent light generation in the all-vdW nanobeam platform.

\begin{figure}[!t]
\centering
\includegraphics[width=0.6\textwidth]{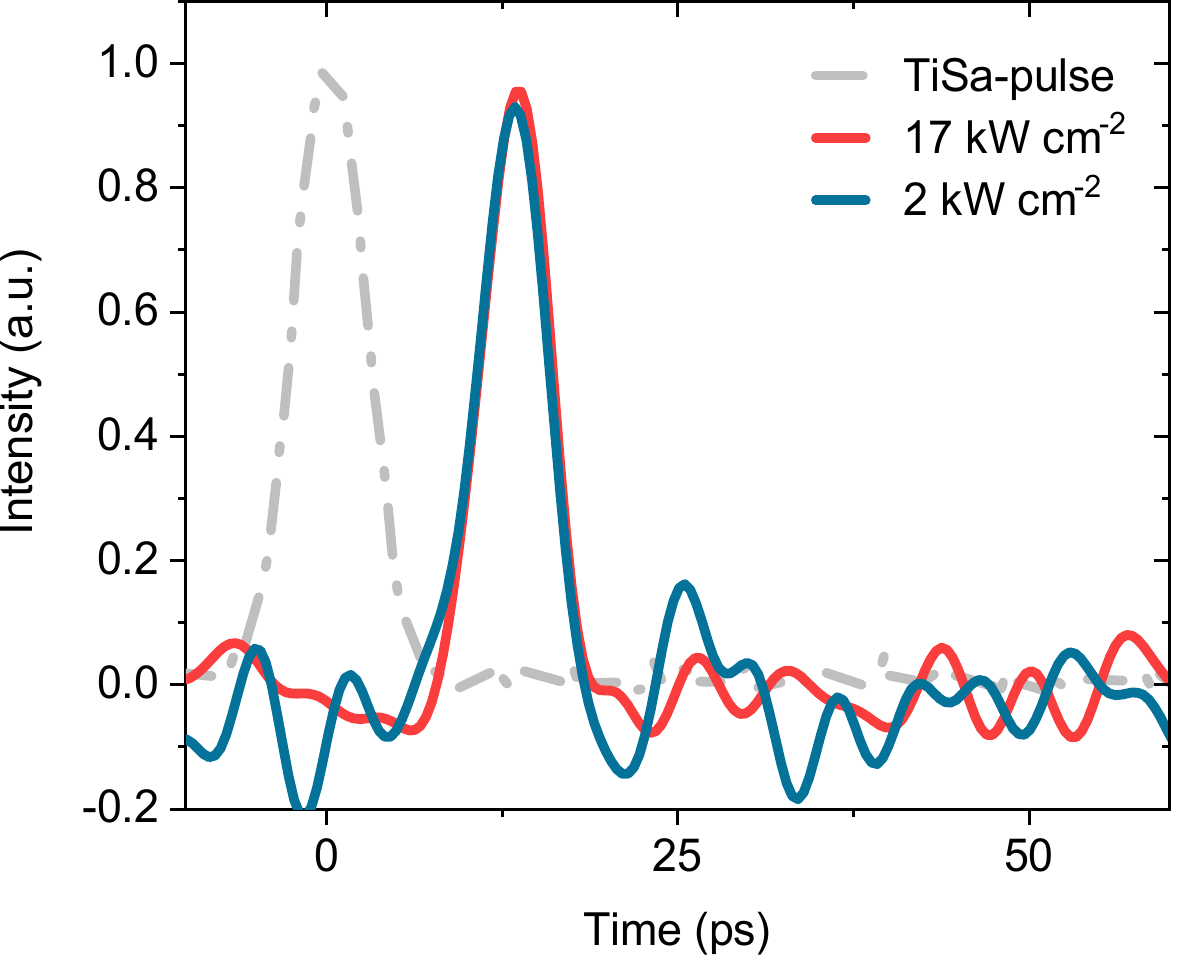}
\caption{\textbf{Results of time-resolved streak-camera measurements of the nanobeam emission.} Temporal response of the pulsed Ti:Sapphire excitation laser (gray dashed curve) and cavity emission recorded below (2~kW\,cm$^{-2}$) and above (17~kW\,cm$^{-2}$) the threshold. The measurements were performed using a synchroscan streak camera with a temporal resolution of approximately 6~ps. The cavity emission remains resolution-limited at both excitation power densities, indicating sub-6-ps emission dynamics and excluding long-lived radiative states or excitonic reservoirs as the origin of the power-dependent broadening observed in the photon-autocorrelation measurements.}
\label{fig:FigureS7}
\end{figure}

\section{Verification of ultrafast dynamics of nanobeam laser in time-resolved measurements with a streak camera}\label{sec:streak} 
To examine whether the excitation-power-dependent broadening of the autocorrelation peaks could originate from changes in the radiative lifetime, time-resolved measurements were performed using a synchroscan streak camera under pulsed excitation. The nanobeam cavities were excited with a mode-locked Ti:Sapphire laser (710~nm, 80~MHz repetition rate, pulse duration $\sim2$~ps). Supplementary Fig.~\ref{fig:FigureS7} compares the temporal response of the excitation laser with the cavity emission recorded below (2~kW\,cm$^{-2}$) and above (17~kW\,cm$^{-2}$) the threshold. The overall temporal resolution of the streak-camera system was approximately 6~ps. At both excitation powers, the cavity emission remains instrument limited and closely follows the temporal profile of the excitation pulse. No measurable temporal broadening or long-lived emission tail is observed within the experimental resolution, indicating that the radiative decay occurs on timescales shorter than 6~ps. These observations are consistent with the intrinsically short radiative lifetime of TMDC ML excitons \cite{Robert2016} and its expected further reduction by cavity-enhanced spontaneous emission. The absence of any detectable slow emission component further excludes long-lived radiative states, trapped excitons or carrier reservoirs as the origin of the excitation-power-dependent broadening of the autocorrelation peaks discussed in the main text. Instead, the observed broadening is consistent with fluctuation-dominated laser dynamics, including delayed pulse build-up and pulse-to-pulse timing jitter, as reported previously for pulsed high-$\beta$ nanolasers operating close to the spontaneous-emission limit \cite{Lebreton2015,Pan2016}.

\section{Temporal broadening and pulse-to-pulse fluctuations in the photon-autocorrelation measurements}\label{sec:broadening}
\begin{figure}[!t]
\centering
\includegraphics[width=\textwidth]{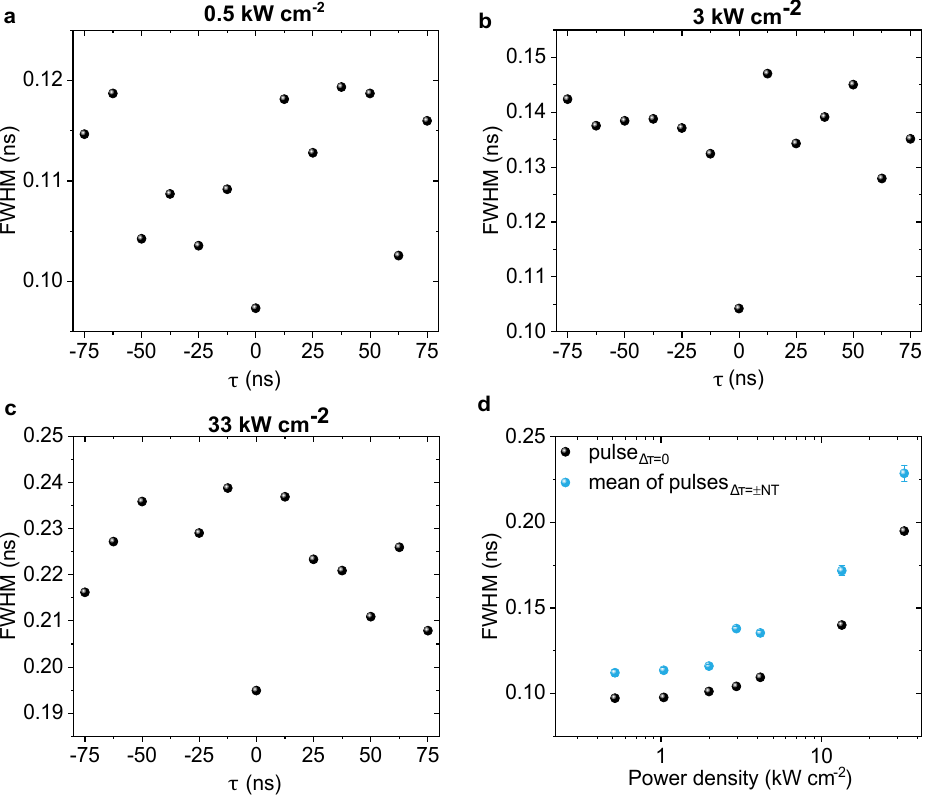}
\caption{\textbf{Temporal broadening of photon-autocorrelation peaks}. \textbf{a-c} FWHM of individual autocorrelation peaks as a function of delay time for excitation power densities of 0.5~kW\,cm$^{-2}$, 3~kW\,cm$^{-2}$, and 33~kW\,cm$^{-2}$, corresponding to operation below, near, and above the nonlinear transition, respectively. \textbf{d} Power-dependent evolution of the FWHM of the central zero-delay peak ($\Delta\tau$=0, black symbols) and the average FWHM of the non-zero-delay peaks ($\Delta\tau=\pm nT$, $n\geq1$, blue symbols). While all correlation peaks broaden with increasing excitation power, the side peaks become systematically broader than the central peak, consistent with the combined influence of nonlinear pulse dynamics and stochastic pulse-to-pulse timing fluctuations.}\label{fig:FigureS8}
\end{figure}

To further investigate the excitation-power-dependent correlation dynamics discussed in the main text, the temporal widths of the individual photon-autocorrelation peaks were analyzed as a function of delay time and excitation power density. Supplementary Fig.~\ref{fig:FigureS8}a--c shows the extracted full width at half maximum (FWHM) of all correlation peaks for representative excitation power densities below threshold (0.5~kW\,cm$^{-2}$), near the nonlinear transition (3~kW\,cm$^{-2}$) and well above threshold (33~kW\,cm$^{-2}$).

Below threshold (Supplementary Fig.~\ref{fig:FigureS8}a), all correlation peaks exhibit nearly identical temporal widths of approximately 0.11~ns, indicating emission dynamics without measurable differences between the zero-delay peak and peaks separated by multiple excitation periods. Close to the nonlinear transition (Supplementary Fig.~\ref{fig:FigureS7}b), the correlation peaks broaden moderately while remaining identical within the experimental uncertainty. In contrast, well above threshold (Supplementary Fig.~\ref{fig:FigureS8}c), all peaks broaden substantially to approximately 0.20--0.24~ns. At the same time, the non-zero-delay peaks become systematically broader than the central autocorrelation peak.

The excitation-power dependence of the extracted peak widths is summarized in Supplementary Fig.~\ref{fig:FigureS8}d. The black symbols denote the FWHM of the central peak ($\Delta\tau=0$), while the blue symbols represent the average FWHM of the side peaks ($\Delta\tau=\pm\,nT$, $n\geq1$). Both quantities increase monotonically with excitation power, demonstrating that the broadening affects interpulse correlations rather than only the central autocorrelation feature. The central peak broadens from approximately 0.10~ns below threshold to about 0.20~ns at the highest excitation powers, whereas the average width of the side peaks reaches approximately 0.23~ns.
These observations indicate two distinct contributions to the temporal broadening. First, the systematic increase of both the central and side-peak widths with excitation power suggests the emergence of intrapulse intensity dynamics associated with stimulated emission. Such behavior is qualitatively consistent with delayed laser build-up and dynamical pulse reshaping previously reported for pulsed high-$\beta$ nanolasers \cite{Pan2016}. Second, the systematic excess broadening of the side peaks indicates an additional stochastic contribution arising from pulse-to-pulse fluctuations of the emission onset time. In this picture, spontaneous-emission-driven timing jitter causes the emission pulse to occur at slightly different times during successive excitation cycles. Photon pairs contributing to the central peak originate from the same excitation pulse and therefore experience only a single timing fluctuation, whereas photons contributing to the side peaks originate from different excitation cycles and accumulate timing uncertainty from two statistically independent lasing events. Consequently, the side peaks are expected to broaden more strongly than the central peak, consistent with previous analyses of stochastic turn-on dynamics in high-$\beta$ nanolasers \cite{Lebreton2015}.

Together, these observations show that the temporal structure of the photon-autocorrelation measurements is governed by a combination of deterministic pulse dynamics and spontaneous-emission-driven timing fluctuations. The systematic excitation-power-dependent broadening therefore provides additional evidence that the nanobeam cavities operate in the fluctuation-dominated high-$\beta$ regime.

\section{Reproducibility of exciton-cavity coupling and high-\texorpdfstring{$\beta$}{beta} lasing in further devices}\label{sec:reproducibility}

\begin{figure}[!t]
\centering
\includegraphics[width=\textwidth]{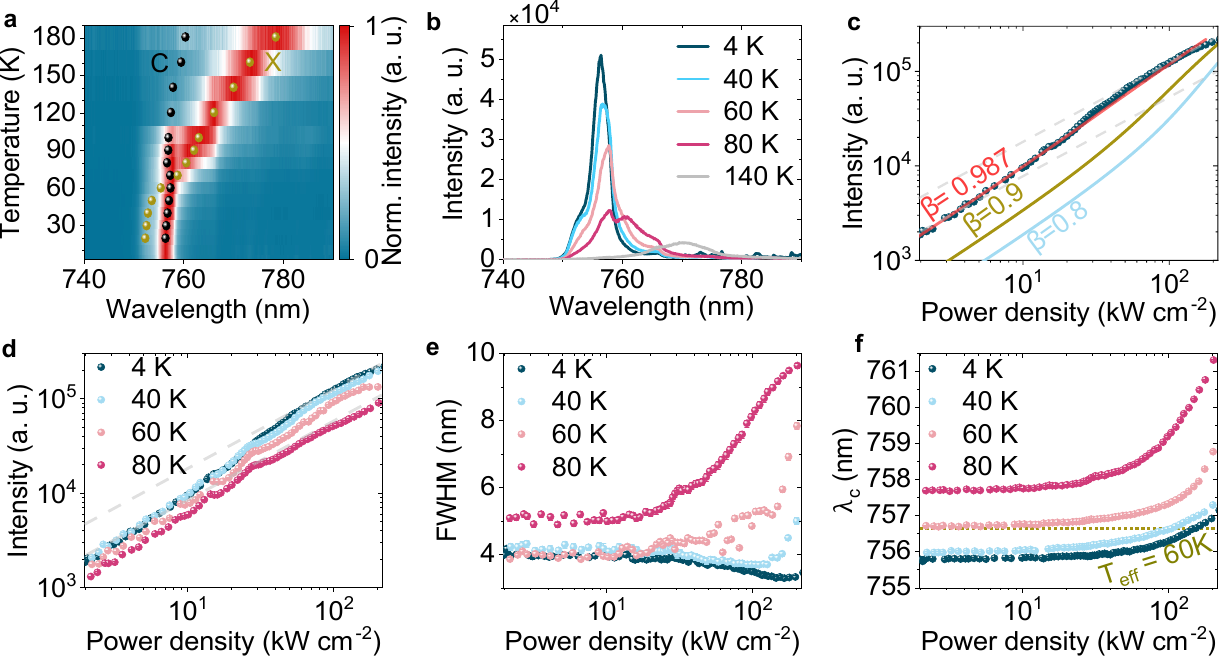}
\caption{\textbf{Reproducibility of exciton-cavity coupling and high-$\beta$ optical signatures in a second all-vdW nanobeam cavity.} \textbf{a} Normalized temperature-dependent \textmu PL map showing the spectral evolution of the cavity mode (C) and MoSe$_2$ exciton (X). The extracted resonance wavelengths are overlaid as symbols. \textbf{b} Representative unnormalized PL spectra recorded at selected temperatures. \textbf{c} I/O characteristics measured at 4~K together with rate-equation fits for different spontaneous-emission coupling factors $\beta$. The best fit yields $\beta=0.987 \pm 0.004$. The gray dashed lines indicate the spontaneous-emission and stimulated-emission regimes. \textbf{d} I/O characteristics measured at temperatures between 4 and 100~K, demonstrating the gradual suppression of the nonlinear transition with increasing temperature. \textbf{e, f} Corresponding excitation-power dependence of the cavity linewidth  and resonance wavelength measured at different temperatures. The increasing linewidth broadening and spectral redshift at elevated temperatures are consistent with enhanced laser-induced heating under optical excitation. The horizontal dotted line indicates an effective temperature $T_{\mathrm{eff}}= 60$~K due to power-dependent laser-induced heating for the nanobeam at $T=4$~K.}\label{fig:FigureS9}
\end{figure}

To assess the reproducibility of the optical properties reported in the main text, a second nanobeam cavity fabricated within the same WS$_2$/MoSe$_2$/WS$_2$ heterostructure was investigated. Supplementary Fig.~\ref{fig:FigureS9} summarizes its temperature- and excitation-power-dependent optical properties.

The temperature-dependent \textmu PL measurements shown in Supplementary Fig.~\ref{fig:FigureS9}a reveal behavior closely resembling that of the device discussed in the main text. The extracted cavity and exciton resonance wavelengths are overlaid on the normalized temperature-dependent emission map, illustrating the pronounced thermal redshift of the MoSe$_2$ exciton and the comparatively weak shift of the cavity mode. Consequently, the exciton and cavity are brought into resonance near $T\approx60$~K. Representative spectra recorded at selected temperatures are shown in Supplementary Fig.~\ref{fig:FigureS9}b. As the excitonic transition approaches the cavity resonance, the cavity emission is strongly enhanced, confirming reproducible exciton-cavity coupling in independently fabricated devices.
The temperature-dependent \textmu PL measurements shown in Supplementary Fig.~\ref{fig:FigureS9}a reveal behavior closely resembling that of the device discussed in the main text. The MoSe$_2$ exciton undergoes the characteristic thermal redshift, whereas the cavity resonance shifts only weakly, enabling continuous tuning of the exciton and cavity mode, which are resonant near $T\approx60$~K. Representative spectra recorded at selected temperatures are presented in Supplementary Fig.~\ref{fig:FigureS9}b. As the exciton approaches the cavity resonance, the cavity emission is strongly enhanced, confirming reproducible exciton-cavity coupling in independently fabricated devices.

The I/O characteristic measured at 4~K is presented in Supplementary Fig.~\ref{fig:FigureS9}c. Similar to the primary device, the cavity exhibits a soft nonlinear transition characteristic of high-$\beta$ lasing. The data were fitted using the rate-equation model described in Section~\ref{sec:rate-equation}. The cavity decay rate was fixed to $\gamma_{\mathrm{cav}}=1.25\times10^{13}$~s$^{-1}$, while the transparency exciton population and dimensionless parameter were estimated as $n_0\approx700$ and $\xi_p=30$, respectively. The fit yields a spontaneous-emission coupling factor of $\beta=(0.987\,\pm\,0.004)$ and a threshold pump density of $P_{\mathrm{th}}=(29\,\pm\,1)$~kW\,cm$^{-2}$, in excellent agreement with the values obtained for the device presented in the main text.

The I/O characteristics measured at different temperatures are summarized in Supplementary Fig.~\ref{fig:FigureS9}d. The pronounced nonlinear transition observed at low temperatures gradually weakens with increasing temperature as the exciton and cavity become spectrally detuned and the optical gain is reduced. The corresponding evolution of the cavity linewidth and resonance wavelength is shown in Supplementary Fig.~\ref{fig:FigureS9}e, f. At low temperatures (4--40~K), the cavity linewidth exhibits slight narrowing accompanied by only a weak pump-induced redshift. At higher temperatures (60--100~K), both the linewidth broadening and the spectral redshift become progressively more pronounced, reflecting increased thermal loading of the nanobeam under optical excitation.

The measured cavity resonance shift furthermore provides an estimate of the laser-induced heating. At a nominal cryostat temperature of 4~K, the cavity wavelength increases from approximately 755.8 nm to 756.9~nm over the investigated excitation range. Comparison with the independently measured temperature-dependent cavity shift indicates an effective cavity temperature of approximately 60~K at the highest pump powers. This heating naturally explains why the strongest nonlinear emission is observed at low nominal cryostat temperatures: optical pumping shifts the operating point towards the temperature at which the exciton and cavity achieve optimal spectral overlap.

Overall, the second nanobeam cavity reproduces all key observations reported in the main text, including temperature-tunable exciton-cavity coupling, soft nonlinear I/O, linewidth narrowing and spontaneous-emission coupling factors approaching unity. These measurements demonstrate that the observed high-$\beta$ lasing behavior is reproducible across the all-vdW nanobeam platform rather than arising from an isolated device.

\bibliography{References}